\documentclass[11pt,a4paper]{llncs}
\usepackage{amssymb,amsmath}
\usepackage{gastex}
\usepackage{fullpage}

\newcommand{\intro}[1]	{\emph{#1}}
\newcommand{\nats}	{\mathbb{N}}
\newcommand{\ints}	{\mathbb{Z}}
\newcommand{\reals}	{\mathbb{R}}
\newcommand{\rationals}	{\mathbb{Q}}
\newcommand{\set}[1]	{\{{#1}\}}

\newcommand{\aut}	{\mathcal{A}}

\newcommand{\gR}	{\mathrel{\mathcal{R}}}
\newcommand{\gL}	{\mathrel{\mathcal{L}}}
\newcommand{\gD}	{\mathrel{\mathcal{D}}}
\newcommand{\gJ}	{\mathrel{\mathcal{J}}}
\newcommand{\gH}	{\mathrel{\mathcal{H}}}

\newcommand{\SD}	{\mathit{SD}}

\newcommand{\cuts}[1]	{\overline{#1}}

\newcommand{\universe}	{\mathcal{U}}
\newcommand{\IFO}	{\interpretation_{\mathit{FO}}}
\newcommand{\IMSO}	{\interpretation_{\mathit{MSO}}}
\newcommand{\LMSO}	{\mathcal{L}_{\mathit{MSO}}}
\newcommand{\cbt}	{\Delta_2}
\newcommand{\structure}	{\mathcal{S}}

\newcommand{\interpretation}	{\mathcal{I}}

\spnewtheorem{fact}[lemma]{Fact}{\bfseries}{\itshape}

\begin{document}

%
%
%

\title{On factorisation forests}
\subtitle{And some applications}
\author{Thomas Colcombet}
\institute{Cnrs/Irisa\\\email{thomas.colcombet@irisa.fr}}
\maketitle

\begin{center}
trees,
	 monadic second-order logic, infinite structures.
\end{center}

\begin{abstract}
The theorem of \emph{factorisation forests} shows the existence of
nested factorisations --- a la Ramsey --- for finite words.
This theorem has important applications in semigroup
theory, and beyond.
The purpose of this paper is to illustrate the importance
of this approach in the context of automata over infinite
words and trees.

We extend the theorem of factorisation
forest in two directions:
we show that it is still valid for any word indexed by
a linear ordering; and we show that it admits a deterministic
variant for words indexed by well-orderings.
A byproduct of this work is also an improvement on the known bounds
for the original result.

We apply the first variant for giving a simplified proof of
the closure under complementation of rational sets
of words indexed by countable scattered linear orderings.
We apply the second variant in the analysis of monadic second-order logic
over trees, yielding new results on monadic interpretations over trees.
Consequences of it are new caracterisations of prefix-recognizable structures
and of the Caucal hierarchy.
\end{abstract}

\section{Introduction}

Factorisation forests were introduced by Simon~\cite{simon90}.
The associated theorem --- which we call the theorem of factorisation forests below ---
states that for every semigroup
morphism from words to a finite semigroup~$S$,
every word has a ramseyan factorisation
tree of height linearly bounded by $|S|$ (see below).
An alternative presentation states that for every morphism~$\varphi$
from~$A^+$ to some finite semigroup~$S$, there exists a
regular expression evaluating to~$A^+$ in which the Kleene exponent
$L^*$ is allowed only when~$\varphi(L)=\set{e}$ for some~$e=e^2\in S$;
i.e. the kleene star is allowed only if it produces
a ramseyan factorisation of the word.

The theorem of factorisation forests
provides a very deep insight on the
structure of finite semigroups, and has therefore
many applications. Let us cite some of them.
Distance automata are nondeterministic finite
automata mapping words to naturals.
An important question concerning them is the limitedness problem:
decide whether this mapping is bounded or not.
It has been shown decidable by Simon
using the theorem of factorisation forests \cite{simon90}.
This theorem also allows a constructive proof of Brown's lemma
on locally finite semigroups \cite{brown71}.
It is also used in the caracterisation of subfamilies
of the regular languages, for instance the polynomial closure of varieties
in~\cite{pinweil97}.
Or to give general caracterisations
of finite semigroups \cite{pinsaecweil91}.
In this last paper, the result is applied
for proving McNaughton's determinisation results of automata
over infinite words \cite{mcnaughton66}.
In the context of languages of infinite words
indexed by~$\omega$, it has also been used
in a complemetation procedure
\cite{LICS06:bojanczyk_colcombet} extending Buchi's lemma~\cite{buchi60}.

The present paper aims first at advertising the
theorem of factorisation forest
which, though already used in many papers, is in fact
known only to a quite limited community.
The reason for this is that all of its proofs
rely on the use of Green's relations: Green's relations
form an extremely important tool in semigroup theory,
but are technical and uncomfortable to work with.
The merit of the factorisation forest theorem
is that it is usable without any significant knowledge
of semigroup theory, while it encapsulates nontrivial
parts of this theory.
Furthermore, as briefly mentionned above and also
in this paper, this theorem as already important
applications to automata theory.
This is why this theorem is worth being advertised
outside the semigroup community as a major tool
in automata theory.

The technical contribution of the paper is an
investigation of the potential use of
factorisation forests in broader contexts than finite words.
An important objective is to be able to apply this theorem
on infinite words, and on trees instead of words.
Those attempts are incarnated by two new variants of the theorem.
As a byproduct we improve the known bounds of the original result
(in particular on the previous improvement \cite{chalopinleung04}).

We also provide some applications of those results.
We give a new proof of the result of Carton and Rispal showing
the closure under comlementation of rational languages of
words with countable scattered linear domain \cite{cartonrispal2005}.
We use the other variant of the theorem
for proving a decomposition result for monadic interpretations
(in fact the application of a technique that we call compaction).
This yields new caracterisations of prefix recognisable
structures and of the Caucal hierarchy.

However, the applications of those results go beyond the one
proposed here.
In paricular, let us mention the work of Blumensath
\cite{blumensath06rabin}
who applies the deterministic variant of the theorem presented
here for giving a new proof of Rabin's theorem \cite{rabin69}.
The theorem of Rabin states that the monadic theory of
the infinite binary tree is decidable.
Different proofs have been proposed for this
result so far, all relying on the use of automata theory, and
most of them on the use of parity games (see \cite{thomas97} for a survey).
For the simpler theory of the naturals with successor
--- originally proved by Buchi \cite{buchi60} ---
another proof technique is known: the compositional method of Shelah
\cite{shelah75}. In this seminal paper, Shelah asks whether there exists a
proof of Rabin's theorem along the same lines.
Blumensath \cite{blumensath06rabin} answers to this
longstanding open question positively.

The content of the paper is organised as follows.
Section~\ref{section:definitions} is  dedicated to
definitions.
Section~\ref{section:factorisation} present
the original theorem of factorisation forests as
well as two less standard presentations of it.
We also introduce in this section the notion
of a ramseyan split, which is central in the remainder
of the paper.
In Section~\ref{section:general-simon} we provide
the first extension of the theorem, the extension to all complete
linear orderings.
Section~\ref{section:scattered} is dedicated to
the application of this extension to the complementation
of automata over countable scattered linear orderings.
In Section~\ref{section:deterministic} we provide the
second extension of the theorem, to ordinals only
this time, but with an extra property of determinism.
Finally, in Section~\ref{section:compaction},
we develop the technique of compaction and use it for
providing a new decomposition result for
monadic interpretations applied to trees. We also show
how this impacts on the theory of infinite structures.

\section{Definitions}
\label{section:definitions}

In this section, we successively present linear orderings,
words indexed by them, semigroups and additive labellings.

\subsection{Linear orderings}
\label{subsection:orderings}

A \intro{linear ordering}~$\alpha=(L,<)$ is a set~$L$
equipped with a total ordering relation~$<$; i.e.
an irreflexive, antisymmetric and transitive relation
such that for every distinct elements~$x,y$ in~~$L$,
either~$x<y$ or~$y<x$.
A \intro{subordering}~$\beta$ of~$\alpha$ is a subset of~$L$ equipped with
the same ordering relation; i.e. $\beta=(L',<)$ with~$L'\subseteq L$.
We write~$\beta\subseteq\alpha$.
We omit the ordering relation~$<$ below unless necessary,
and just say that~$L$ is a linear ordering.
An \intro{convex subset of~$\alpha$} is a subset~$S$ of~$\alpha$
such that for all~$x,y\in S$ and~$x<z<y$, $z\in S$.
We use the notations
$[x,y],[x,y[,]x,y],]x,y[,]-\infty,y],]-\infty,y[,
[x,+\infty[$ and~$]x,+\infty[$
for denoting the usual \intro{intervals}.
Intervals are convex,
but the converse does not hold in general.
Given two subsets~$X,Y$ of a linear ordering,
$X<Y$ holds if for all~$x\in X$ and~$y\in Y$, $x<y$.

The \intro{sum} of two linear orderings~$\alpha_1=(L_1,<_1)$ and
$\alpha_2=(L_2,<_2)$ (up to renaming, assume~$L_1$ and~$L_2$ disjoint),
denoted~$\alpha_1+\alpha_2$,
is the linear ordering~$(L_1\cup L_2,<)$ with~$<$
coinciding to~$<_1$ on~$L_1$, to~$<_2$ on~$L_2$
and such that~$L_1<L_2$.
More generally, given a linear ordering $\alpha=(L,<)$
and for each~$x\in L$ a linear ordering~$\beta_x=(K_x,<_x)$
(the $K_x$ are assumed disjoint),
we denote by~$\sum_{x\in\alpha}\beta_x$
the linear $(\cup_{x\in L} K_x,<')$
with~$x'<'y'$ if~$x<y$ or~$x=y$ and~$x'<_xy'$,
where~$x'\in K_x$ and~$y'\in K_y$.

A linear ordering~$\alpha$ is \intro{well ordered} if
every nonempty subset has a minimal element.
It is \intro{complete}
if every nonempty subset of~$\alpha$ with an upper bound
has a least upper bound in~$\alpha$,
and every nonempty subset of~$\alpha$ with a lower bound
has a greatest lower bound in~$\alpha$.

A \intro{cut} in a linear ordering~$\alpha=(L,<)$
is a couple~$(E,F)$ where~$\set{E,F}$ is a partition of~$L$,
and~$E<F$.
Cuts are totally ordered by~$(E,F)<(E',F')$ if~$E\subsetneq E'$.
This order has a minimal element~$\bot=(\emptyset,L)$
and a maximal element~$\top=(L,\emptyset)$. We denote by~$\cuts\alpha$
the set of cuts over~$L$ and by $\cuts\alpha^*$
the set~$\cuts \alpha\setminus\set{\bot,\top}$.
An important remark is that $\cuts\alpha$
and~$\cuts\alpha^*$ are complete linear orderings.

Cuts can be thought as new elements located between the elements of~$L$:
given~$x\in L$, $x^-=(]-\infty,x[,[x,+\infty[)$
represents the cut placed just before~$x$, while
$x^+=(]-\infty,x],]x,+\infty[)$ is the cut placed just after~$x$.
We say in this case that~$x^+$ is \intro{the successor of~$x^-$ through~$x$}.
But not all cuts are successors or predecessors of another cut.
A cut~$c$ is a \intro{right limit} (resp. a \intro{left limit})
if it is not the minimal element and
not of the form~$x^+$ for some~$x$ in~$L$
(resp. not the maximal element and not of the form~$x^-$).

Two linear orderings~$\alpha=(L,<)$ and~$\beta=(L',<')$
are \intro{isomorphic} if there exists a bijection~$f$ from~$L$
onto~$L'$ such that for every~$x,y$ in~$L$, $x<y$ iff
$f(x)<'f(y)$. In this case, we also say that~$(L,<)$
and~$(L',<')$ have \intro{the same order type}. This is an equivalence
relation on the class of linear orderings.
We denote by~$\omega,\omega^*,\zeta$ the order types of
respectively~$(\nats,<)$ (the naturals), $(-\nats,<)$ (the nonpositive
integers) and~$(\ints,<)$ (the integers).
The order type of a well-ordering is called an~\intro{ordinal}.
Below, we do often not distinguish between a linear ordering and its type.
This is safe since all the construction we perform are isomorphism invariant.

The interested reader can find in~\cite{Rosenstein82}
additional material on linear orderings.

\subsection{Words}
\label{subsection:words}

We use a generalized version of words: words
indexed by a linear ordering.
Given a linear ordering~$\alpha=(L,<)$ and a finite alphabet~$A$,
an \intro{$\alpha$-word~$u$ over the alphabet~$A$}
is a mapping from~$L$ to~$A$.
We also say that~$\alpha$ is the \intro{domain} of the word~$u$,
or that~$u$ is a word \intro{indexed} by~$\alpha$.
Standard finite words are simply the words indexed
by finite linear orderings.
Given a word~$u$ of domain~$\alpha$ and~$\beta\subseteq\alpha$,
we denote by~$u|_\beta$ the word~$u$ restricted to its positions
in~$\beta$.

Given an $\alpha$-word~$u$ and a~$\beta$-word~$v$,
$uv$ represents the $(\alpha+\beta)$-word
defined by~$(uv)(x)$ is $u(x)$ if~$x$ belongs to~$\alpha$
and~$v(x)$ if~$x$ belongs to~$\beta$. This construction is
naturally generalized to the infinite product
$\prod_{i\in\alpha}u_i$, where~$\alpha$ is an order type
and~$u_i$ are linear $\beta_i$-words; the resulting
being a~$\sum_{i\in\alpha}\beta_i$-word.

\subsection{Semigroups and additive labellings}
\label{subsection:semigroups}

For a thorough introduction to semigroups, we refer the reader to
\cite{lallement79,pinvariete84,pinvariety86}.
A \intro{semigroup} $(S,.)$ is a set~$S$ equipped
with an associative binary operator written multiplicatively.
Groups and monoids are particular instances of semigroups.
The set of nonempty finite words~$A^+$
over an alphabet~$A$ is a semigroup -- it is the semigroup
freely generated by~$A$.
A \intro{morphism of semigroups} from a semigroup~$(S,.)$
to a semigroup~$(S',.')$ is a mapping~$\varphi$ from~$S$ to~$S'$
such that for all~$x,y$ in~$S$, $\varphi(x.y)=\varphi(x).'\varphi(y)$.
An \intro{idempotent} in a semigroup is an element~$e$ such that~$e^2=e$.

Let~$\alpha$ be a linear ordering and~$(S,.)$
be a semigroup.
A mapping~$\sigma$
from couples $(x,y)$ with $x,y\in\alpha$ and $x<y$ to~$S$
is called an \intro{additive labelling} if
for every~$x<y<z$ in~$\alpha$,
$\sigma(x,y).\sigma(y,z)=\sigma(x,z)$.

Given a semigroup morphism~$\varphi$ from~$(A^\diamond,.)$
to some semigroup~$(S,.)$
and a word~$u$ in~$A^\diamond$ of domain~$\alpha$,
there is a natural way to construct
an additive labelling~$\phi_u$ from~$\cuts\alpha$
to~$(S,.)$: for every two cuts~$x<y$ in~$\cuts\alpha$,
set~$\varphi_u(x,y)$ is~$\varphi(u|_{]x,y[})$.
I.e. $\varphi_u(x,y)$ is the image by~$\varphi$
of the factor of~$u$ located between~$x$ and~$y$.
We denote by~$\varphi^*_u$ the mapping~$\varphi_u$ restricted to~$\cuts \alpha^*$.

\subsection{Structures, graphs, trees, logics}
\label{subsection:definitions-logics}

\subsubsection*{Relational structures}

Let us first remark that the definitions presented here are useless
before Section~\ref{section:deterministic}, have marginal
consequences in Section~\ref{section:deterministic},
and are of real interest only for Section~\ref{section:compaction}.

A \intro{relational structure} $(\universe,R_1,\dots,R_n)$
is a set~$\universe$, called the \intro{universe}, together with
\intro{relations} $R_1,\dots,R_n$ of fixed finite arity over~$\universe$.
Each relation~$R$ has a \intro{name} that we write~$R$ itself.
The \intro{signature} of a structure contains the names involved together
their arity.
A \intro{graph} is a relational structure for which the relations have arity
$1$ and one relation of arity~$2$.
The elements of the universe are called \intro{vertices},
the unary relations are called \intro{label relations},
and the binary relations is called the \intro{edge relation}.
A \intro{path} is a finite sequence of vertices
such that two successive vertices are in relation by
the edge relation. The first vertex is called the \intro{origin} of
the path, and the last vertex the \intro{destination}.

Linear orderings can be naturally represented as graphs:
$(L,<)$ can be seen as a graph of vertices~$L$,
with an edge between~$x$ and~$y$ iff~$x<y$.
For a linear ordering~$\alpha=(L,<)$
and a finite alphabet~$A=\{a_1,\dots,a_n\}$,
an $\alpha$-word $u$ is the graph $(L,<,a_1,\dots,a_n)$
obtained from the graph of the linear ordering
by setting~$a_i$ to be interpreted as $u^{-1}(a_i)$;
the set of positions in the word corresponding to letter~$a_i$.

A \intro{tree} $t$ is a graph such that there is only one edge relation,
called the \intro{ancestor relation}
and denoted $\sqsubseteq$, satisfying:
\begin{itemize}
\item the relation $\sqsubseteq$ is an order,
\item there is a minimal element for~$\sqsubseteq$, called the \intro{root},
\item for every~$u$, the set $\{v~:~v\sqsubseteq u\}$ is an ordinal
	of length at most~$\omega$.
\end{itemize}
The vertices of a tree are called \intro{nodes}.
Maximal chains of nodes in a tree are called \intro{branches}.

\medskip

{\bf Warning:}
The trees are \emph{not} defined by a `direct successor' relation, but
rather by the ancestor relation. This has major impact on the logical side:
all the logics we use below can refer to the ancestor relation,
and it is well-known that first-order logic using this ancestor relation is
significantly more expressive over trees than first-order logic with
access to the successor of a node only. The results would fail if the
ancestor relation was not used.

\medskip

A particular tree will play a special role below. The \intro{complete binary tree}
has as universe $\{0,1\}^*$, as ancestor relation the prefix relation,
and has two unary relations, $0=\{0,1\}^*0$ and~$1=\{0,1\}^*1$.
We call the relation~$0$ the \intro{left-child relation},
while $1$ is the \intro{right-child relation}.
We denote by~$\cbt$ the complete binary tree.

One constructs a tree from a graph by unfolding.
Given a graph $G$ and one of its vertices~$v$,
the \intro{unfolding} of~$G$ from~$v$ is the tree which has
as nodes the all paths with origin~$v$, as ancestor relation the
prefix relation over paths, and such that a path~$\pi$ is
labelled by~$a$ in the unfolding
iff its destination is labelled by~$a$ in the graph.

\subsubsection*{Logics}

For defining first-order logic, we need to have at our disposal
a countable set of \intro{first-order variables} $x,y,\dots$
to pick from.
The \intro{atomic formul\ae} are~$R(x_1,\dots,x_n)$
for~$x_1,\dots,x_n$ first-order variables
and~$R$ the name of relation of arity~$n$;
given two first-order variables $x,y$, $x=y$ is also an
atomic formula.
\intro{First-order} logic formul\ae{} are made out of
these atomic formul\ae, combined with the boolean connectives
$\vee,\wedge,\neg$, and the first-order quantifiers~$\exists x$ and
$\forall x$.
For monadic logic, we need furthermore a countable set of \intro{monadic
variables} $X,Y,\dots$
\intro{Monadic (second-order) formul\ae} are defined as first-order formul\ae{},
but further allow the use of monadic quantifiers $\exists X$, $\forall X$,
and of a membership atomic formula $x\in X$, where~$x$ is a first-order
variable and~$X$ a monadic one.
For first-order as well as monadic formul\ae{} we use the standard notion
of \intro{free variables}.
A formula without free variables is called a \intro{closed} formula.

We denote by~$\structure\models\phi$ the fact, for a closed formula~$\phi	$
and a structure~$\structure$, that the formula is true over the structure~$\structure$.
The formal definition uses the standard semantic, the value of first-order variables
ranging over elements of the universe of the structure, while monadic variables take as
values subsets of the universe.
We say that~$\structure$ is a \intro{model} of~$\phi$, or that
$\phi$ is \intro{satisfied} over~$\structure$. When the structure
is obvious from the context, we simply state that $\phi$ is satisfied.
We also allow ourselves to use formul\ae{} like~$\phi(x_1,\dots,x_n)$ to denote
that the free-variables of~$\phi$ are among~$\{x_1,\dots,x_n\}$.
Then given elements~$u_1,\dots,u_n$ in the universe of a structure~$\structure$,
we write $\structure\models\phi(u_1,\dots,u_n)$ if the formula~$\phi$
is true over the structure~$\structure$, using the valuation which to
each $x_i$ associates $u_i$.

A relational structure~$\structure$ has a
\intro{decidable $L$-theory} (where $L$ is either first-order or monadic),
if there is an algorithm which, given a formula~$\phi$ of the logic~$L$,
answers whether~$\structure\models\phi$ or not.

\subsubsection*{Interpretations}

An interpretation is an operation defined by logic formul\ae{} that defines
a structure inside another one.
An \intro{interpretation} is given as a tuple
\begin{align*}
\interpretation&=
(\delta(x),\phi_1(x_1,\dots,x_{|R_1|}),\dots,\phi_k(x_1,\dots,x_{|R_K|}))
\end{align*}
where~$\delta(x),\phi_1(x_1,\dots,x_{|R_1|}),\dots,\phi_k(x_1,\dots,x_{|R_K|})$
are formul\ae{} of corresponding free variables.
The interpretation is \intro{first-order} if the formul\ae{} are first-order and
\intro{monadic} if the formul\ae{} are monadic.

Given a structure~$\structure$ of universe~$\universe$,
$\interpretation(\structure)$ is the structure of universe
\begin{align*}
\universe_{\interpretation(\structure)}&=\{u\in\universe~:~\structure\models\delta(u)\}\ ,
\end{align*}
and such that the interpretation of~$R_i$ is
\begin{align*}
\{(u_1,\dots,u_{|R_i|})\in\universe_{\interpretation(\structure)}^{|R_i|}
~:~\structure\models\phi_i(u_1,\dots,u_{|R_i|})\}.
\end{align*}

A special case of interpretation is the \intro{marking}.
A marking replicates the structure, and adds some new unary relations
on it.

\section{Factorisation forest theorem: various presentations for the standard case}
\label{section:factorisation}

In this section, we present the theorem of factorisation forest.
We first give the original statement in Section~\ref{subsection:simon-standard}.
The in Section~\ref{subsection:simon-regular}, we provide another equivalent
presentation in terms of regular expressions; possibly the most natural one.
In Section~\ref{subsection:simon-split}, we introduce the notion of
a split, and use it for a third formalisation of the result.
This notion is the one used in the extensions
of the factorisation forest theorem we provide below.

\subsection{Factorisation forest theorem}
\label{subsection:simon-standard}

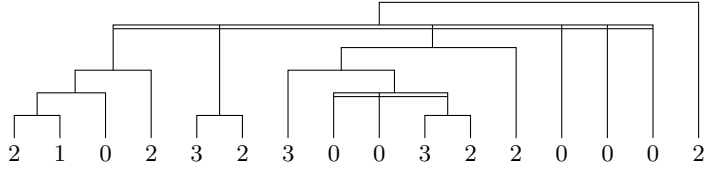
\begin{figure}[ht]
\begin{center}
\begin{picture}(110,25)(-55,-2)

  \gasset{Nadjust=w,Nadjustdist=2,Nh=6,Nmr=1}

  \node[Nframe=n](G1)(-48,0){$2$}
  \node[Nframe=n](G2)(-42,0){$1$}
  \node[Nframe=n](G3)(-36,0){$0$}
  \node[Nframe=n](G4)(-30,0){$2$}
  \node[Nframe=n](G5)(-24,0){$3$}
  \node[Nframe=n](G6)(-18,0){$2$}
  \node[Nframe=n](G7)(-12,0){$3$}
  \node[Nframe=n](G8)(-6,0){$0$}
  \node[Nframe=n](G9)(-0,0){$0$}
  \node[Nframe=n](G10)(6,0){$3$}
  \node[Nframe=n](G11)(12,0){$2$}
  \node[Nframe=n](G12)(18,0){$2$}
  \node[Nframe=n](G13)(24,0){$0$}
  \node[Nframe=n](G14)(30,0){$0$}
  \node[Nframe=n](G15)(36,0){$0$}
  \node[Nframe=n](G16)(42,0){$2$}

  \gasset{AHnb=0}

  \drawline(-48,2)(-48,5)(-42,5)(-42,2)
  \drawline(-45,5)(-45,8)(-36,8)(-36,2)
  \drawline(-40,8)(-40,11)(-30,11)(-30,2)

  \drawline(-24,2)(-24,5)(-18,5)(-18,2)

  \drawline(6,2)(6,5)(12,5)(12,2)
  \drawline(-6,2)(-6,8)(9,8)(9,5)
  \drawline(-6,7.5)(9,7.5)
  \drawline(0,2)(0,8)
  \drawline(-12,2)(-12,11)(2,11)(2,8)
  \drawline(-5,11)(-5,14)(18,14)(18,2)

  \drawline(-35,11)(-35,17)(36,17)(36,2)
  \drawline(-35,16.5)(36,16.5)
  \drawline(-21,5)(-21,17)
  \drawline(7,14)(7,17)
  \drawline(24,2)(24,17)
  \drawline(30,2)(30,17)

  \drawline(0,17)(0,20)(42,20)(42,2)
   \end{picture}
\end{center}

\caption{A factorisation tree}
\label{figure:factorisation-tree}
\end{figure}

Fix an alphabet~$A$ and a semigroup morphism~$\varphi$
from~$A^+$ to a finite semigroup~$(S,.)$.
A \intro{factorisation tree} of a word~$u\in A^+$
is an ordered unranked tree in which
each node is either a leaf labeled by a letter,
or an internal node, and such that
the word obtained by reading the leaves from
left to right (the yield) is~$u$.
The \intro{height} of the tree is defined as usual, with the convention that
the height of a tree restricted to a single leaf is $0$.
A factorisation tree is \intro{ramseyan} (for~$\varphi$)
if every node 1) is a leaf, or 2) has two children, or,
3) the values of its children are all mapped by~$\varphi$
	to the same idempotent of~$S$.

\begin{example}
Fix $A=\set{0,1,2,3,4}$,
$(S,.)=(\ints/5\ints,+)$ and~$\varphi$
to be the only semigroup morphism from~$A^+$ to~$(S,.)$
mapping each letter to its value.
Figure~\ref{figure:factorisation-tree} presents
a ramseyan factorisation tree for the word $u=210232300322002$
($u$ is the yield of the tree).
In this drawing, internal nodes appear as horizontal lines.
Double line correspond to case~3 in the description of
ramseyanity.
\end{example}

The theorem of factorisation forests is then the following.
\begin{theorem}[factorisation forests]
\label{theorem:simon}
For every alphabet~$A$, finite semigroup~$(S,.)$,
semigroup morphism~$\varphi$ from~$A^+$ to~$S$ and word~$u$ in~$A^+$,
$u$ has a ramseyan factorisation tree of height at most~$3|S|$.
\end{theorem}
The original theorem is due to Simon \cite{simon90},
with a bound of $9|S|$. An improved bound of $7|S|$
is provided by Chalopin and Leung \cite{chalopinleung04}.
The value of $3|S|$ is a byproduct of the present work.

\subsection{A variant via regular expressions}
\label{subsection:simon-regular}

The use of factorisation trees gave the name of factorisation forests to the theorem.
But it is sometime very convenient to use another formalisation in
terms of regular expressions.
This presentation is new (to the knowledge of the author),
but its simplicity makes it worth to be advertised.
Let~$A$ be an alphabet, $\varphi$ a semigroup morphism
from~$A^+$ to some semigroup~$S$, and $E$ be a regular expression
over the alphabet~$A$. $E$ is \intro{$\varphi$-ramseyan}
if for each occurence~$L^*$ of the Kleene star in~$E$,
$L$ is mapped to~$\{e\}$ by~$\varphi$, for~$e$ an idempotent in~$S$.
\begin{example}
Let~$S$ be $\ints/2\ints$ with the addition, $A$ be~$\{0,1\}$ and
$\varphi$ be the morphism from~$A^+$ to~$S$ sending each letter to
its value modulo $2$.
The expression $0(0+10^*1)^*+10^*1(0+10^*1)^*$ is $\varphi$-ramseyan
and evaluates to~$\varphi^{-1}(0)$.
\end{example}

\begin{theorem}[variant of factorisation forests]\label{theorem:simon-regular}
For every alphabet~$A$, finite semigroup~$(S,.)$,
semigroup morphism~$\varphi$ from~$A^+$ to~$S$ and $x$ in~$S$,
there exists a $\varphi$-ramseyan
regular expression $E_x$ evaluating to~$\varphi^{-1}(x)$.
\end{theorem}
\begin{proof}
By induction on~$k$, for every~$x$ in~$S$, let
the $\varphi$-ramseyan regular expression~$E^{k}_{x}$ be:
\begin{align*}
E^{0}_x   & = \varphi^{-1}(x)\cap A\ , &
E^{k+1}_x & = E_x^{k} + \sum_{yz=x} E_y^{k} E_{z}^{k}
+ \sum_{e^2=e=x} (E_e^{k})^*\ .
\end{align*}
On can show by induction on~$k$ that
for all~$x\in S$, $E^k_x$ evaluates to the set of words in~$\varphi^{-1}(x)$
possessing a factorisation tree of height $k$.
This proof, for both directions of the inclusion, is a direct application
of the definitions.
Then, by Theorem~\ref{theorem:simon}, $E^{3|S|}_x$ evaluates to $\varphi^{-1}(x)$.\qed
\end{proof}

The interest of Theorem~\ref{theorem:simon-regular}
is that it allows to perform proofs
by induction on the structure of ramseyan regular expressions.
By the following refinement, we can derive complexities when using this technique.
\begin{property}[refinement of Theorem~\ref{theorem:simon-regular}]\label{property:bounds-regular-simon}
The height of the regular expression $E_x$ is at most $3|S|+1$, counting $0$
for the operator~$+$, and $1$ for the concatenation, the Kleene star and
constants.
The regular expression $E_x$ contains at most $6|S|^2$ distinct subexpressions,
at most $3|S|^2$ distinct subexpressions
without the $+$-operator at the root.
\end{property}
Those bounds are obtained from the last variant, Theorem~\ref{theorem:simon-split}.

\subsection{A variant via ramseyan splits}
\label{subsection:simon-split}

The third equivalent presentation to the theorem
of factorisation forests uses the notion of ramseyan splits.
One way to see a split is as a form of presentation of a tree.
This formalisation naturally extends to infinite words,
and is very natural to use in automata theoretic constructions.
The extensions of the theorem proposed in the remaining of
the paper use this definition.

A \intro{split of height~$N$} of a linear ordering~$\alpha$
is a mapping~$s$ from~$\alpha$ to~$[1,N]$.
Given a split, two elements $x$ and~$y$ in~$\alpha$
such that~$s(x)=s(y)=k$ are \intro{$k$-neighbours}
if $s(z)\geq k$ for all~$z\in[x,y]$.
$k$-neighbourhood is an equivalence relation over~$s^{-1}(k)$.
Fix an \intro{additive labelling} from~$\alpha$
to some finite semigroup~$S$.
A split of~$\alpha$ is \intro{ramseyan} for~$\sigma$
--- we also say a \intro{ramseyan split for~$(\alpha,\sigma)$} ---
if for every~$k\in[1,N]$, every~$x<y$
and~$x'<y'$ such that all the elements~$x,y,x',y'$ are $k$-neighbours,
then~$\sigma(x,y)=\sigma(x',y')=(\sigma(x,y))^2$;
Equivalently, for all~$k$, every class of~$k$-neighbourhood
is mapped by~$\sigma$ to a single idempotent of the semigroup.

\begin{example}\label{example:split}
Let~$S$ be~$\ints/5\ints$ equipped with the addition~$+$.
Consider the linear ordering of~$17$ elements and the additive labelling~$\sigma$
defined by:
\begin{align*}
\begin{array}{c c c c c c c c c c c c c c c c c c c c c c c c c c c c c c c c c}
	      |&3&|&1&|&0&|&2&|&3&|&2&|&3&|&0&|&0&|&3&|&2&|&2&|&0&|&0&|&0&|&2&|\\
\end{array}
\end{align*}
Each symbol `$|$' represents an element, the elements being ordered from left to right.
Between two consecutive elements~$x$ and~$y$ is represented the value of~$\sigma(x,y)\in S$.
In this situation, the value of~$\sigma(x,y)$
for every~$x<y$ is uniquely defined according to the additivity
of~$\sigma$: it is obtained by summing all the values between~$x$ and~$y$
modulo~$5$.

A split~$s$ of height~$3$ is the following, where we have written above each element~$x$
the value of~$s(x)$:
\begin{align*}
\begin{array}{c c c c c c c c c c c c c c c c c c c c c c c c c c c c c c c c c}
	      1& &3& &2& &2& &1& &2& &1& &2& &2& &2& &3& &2& &1& &1& &1& &1& &2\\
	      |&2&|&1&|&0&|&2&|&3&|&2&|&3&|&0&|&0&|&3&|&2&|&2&|&0&|&0&|&0&|&2&|
\end{array}
\end{align*}
In particular, if you choose~$x<y$ such that~$s(x)=s(y)=1$, then the sum
of elements between them is~$0$ modulo~$5$.
If you choose~$x<y$ such that~$s(x)=s(y)=2$ but there is no element~$z$ in between
with~$s(z)=1$ --- i.e. $x$ and~$y$ are~$2$-neighbours --- the sum of values
separating them is also~$0$ modulo~$5$.
Finally, it is impossible to find two distinct $3$-neighbours in our example.
\end{example}

\begin{theorem}\label{theorem:simon-split}
For every finite linear ordering $\alpha$, every finite
semigroup~$(S,.)$ and additive labelling~$\sigma$ from~$\alpha$
to~$S$, there exists a ramseyan  split for~$\alpha$
of height at most~$|S|$.
\end{theorem}
The proof of this result is postponed to Section~\ref{subsection:simon-general-proof},
as the proof is a simplification of the proof of its extension
Theorem~\ref{theorem:general-simon}.

Let us state the link between ramseyan  splits and factorisation trees.
Fix an alphabet~$A$, a semigroup~$S$, a morphism~$\varphi$ from~$A^+$ to~$S$
and a word~$u\in A^+$. The following is easy to establish:
\begin{itemize}
\item every ramseyan factorisation tree of height~$k$ of~$u$
	can be turned into a ramseyan  split of height at most~$k$ of~$\varphi^*_u$,
\item every ramseyan  split of height~$k$ of~$\varphi^*_u$ can be turned
	into a factorisation tree of height at most~$3k$ of~$u$.
\end{itemize}
Using this last argument and Theorem~\ref{theorem:simon-split},
we directly obtain a proof of Theorem~\ref{theorem:simon}
with the announced bound of $3|S|$.
Using similar arguments, one obtains the bounds of
Property~\ref{property:bounds-regular-simon}.

\section{Extension of the factorisation forest theorem to infinite words}
\label{section:general-simon}

The contribution of this section is an extension of Theorem~\ref{theorem:simon-split} to complete linear orderings.
\begin{theorem}\label{theorem:general-simon}
For every complete linear ordering~$\alpha$, every finite
semigroup~$(S,.)$ and additive labelling~$\sigma$ from~$\alpha$
to~$S$, there exists a ramseyan  split for~$(\alpha,\sigma)$
of height at most~$3|S|$ ($|S|$
if~$\alpha$ is an ordinal).
\end{theorem}
Compared to Theorem~\ref{theorem:simon-split},
we trade the finiteness --- which is replaced by the completeness --- for a bound of $3|S|$
--- which replaces a bound of $|S|$. The special case of~$\alpha$ being an ordinal,
proves Theorem~\ref{theorem:simon-split}.

The remaining of the section is devoted to the proof of Theorem~\ref{theorem:general-simon},
as well as its ordinal version, Theorem~\ref{theorem:simon-split}.
We start in Section~\ref{subsection:linear-orders} by establishing
some elementary topological lemmas relative to complete linear orderings.
Then, in Section~\ref{subsection:simon-general-proof},
we give successively a proof of both Theorems~\ref{theorem:simon-split} and~\ref{theorem:general-simon}.

\subsection{On linear orderings}
\label{subsection:linear-orders}

The subject of this section is to provide preparatory lemmas
on linear orderings. Namely Lemmas~\ref{lemma:firstS} and~\ref{lemma:modulo}.
This Section is not relevant for the simpler proof of Theorem~\ref{theorem:simon-split}.

We consider here a binary relation~$R$
over a linear ordering~$\alpha$.
The statement~$R(x,y)$ can be thought as meaningful only for
$x<y$, in the sense that we do not take into account the value
of $R$ elsewhere.
We say that a binary relation~$R$ over~$\alpha$ is \intro{upward closed}
if for every~$x\leq x'<y'\leq y$, $R(x',y')$ implies~$R(x,y)$.

\begin{lemma}\label{lemma:firstS}
Let~$\alpha$ be a complete linear ordering,
and~$R$ be an upward closed relation over~$\alpha$.
There exists~$\gamma\subseteq\alpha$ such that for every~$x<y$ in~$\alpha$,
\begin{itemize}
\item if~$R(x,y)$ then $[x,y]\cap \gamma$ is nonempty,
\item if~$]x,y[\cap \gamma$ contains two distinct elements, then $R(x,y)$.
\end{itemize}
\end{lemma}

Let us first remark that if Lemma~\ref{lemma:firstS}
holds for some linear ordering~$\alpha$, then it is also true
for every convex subset of~$\alpha$.
For this reason, we can safely add a new minimal element~$\bot'$ and maximal
element~$\top'$ to~$\alpha$, such that for every~$x$ in~$\alpha$,
$R(\bot',x)$ and~$R(x,\top')$.
Define now for~$x\in \alpha$,
\begin{align*}
		l(x)&=\sup\set{y\,:\,\forall z>x.~R(y,z)}\ ,\\
\text{and}\quad r(x)&=\inf\set{z\,:\,\forall y<x.~R(y,z)}\ .
\end{align*}
Thanks to the adjunction of~$\bot'$ and~$\top'$, $l$ and~$r$ are defined
everywhere but for the minimal and maximal elements respectively.

\begin{fact}\label{fact:proplr} The following holds.
\begin{enumerate}
\item Both~$l$ and~$r$ are nondecreasing.
\item For every~$x$, $l(x)\leq x\leq r(x)$.
\item For every~$x$, $l(x)=x$ iff~$r(x)=x$.
\item For every~$x$, $r^\omega(x)=\sup\set{r^n(x):n\in\nats}$ and~$l^\omega(x)=\inf\set{l^n(x):n\in\nats}$ are fixpoints
	of both~$l$ and~$r$.
\item For every~$x,y,z$, if~$x<z\leq r(z)<y$ then~$R(x,y)$.
\item For every~$x,y,z$, if~$z<x<y<r(z)$ then~$\neg R(x,y)$.
\end{enumerate}
\end{fact}
\begin{proof}
Items 1,2,5 and 6 follow from the definition.

For item 3. By upward closure of~$R$,
$l(x)=x$ iff for every~$y<x$ and~$z>x$, $R(y,z)$, iff $r(x)=x$.

For item 4. Let~$y=r^\omega(x)$.
By item 2, we have~$y\leq r(y)$.
We have to prove~$r(y)\leq y$.
Let~$x_n$ be~$r^n(x)$.
If~$x_{n+1}=x_n$ for some~$n$, then~$y=x_n=r(x_n)=r(y)$.
Else~$x_0<x_1<\dots<y$.
It follows by definition of~$r$ that for all~$n$, $R(x_n,y)$.
This implies~$r(y)=y$.\qed
\end{proof}

\newcommand{\Fix}{\mathrm{Fix}}

We can now prove Lemma~\ref{lemma:firstS}.
\begin{proof}
Set~$\Fix$ to be the set of fixpoints of~$r$ (equivalently, $l$).
Define the equivalence relation~$\sim$ by~$x\sim y$ if
$x=y\in\Fix$ or~$[x,y]\cap\Fix$ is empty.
This relation induces two kind of equivalence classes:
singletons consisting of a single fixpoint, or maximal intervals containing
no fixpoint.

Let~$C$ be an equivalence class of~$\sim$.
If~$C=\set{x}$ for~$x\in\Fix$, set~$\gamma(C)$ to be~$C$.
Else, $C$ is an interval.
Fix an element~$x_C$ in~$C$, set~$x_C^n$ to be~$r^n(x_C)$ for~$n\geq 0$ and~$x_C^{-n}$
be~$l^n(x)$ for~$n\geq 0$ (both definitions coincide for~$n=0$ with~$x_C^0=x_C$).
By induction and using fact~\ref{fact:proplr}, one easily shows that
for every~$n$, both~$x_C^n$ and~$x_C^{-n}$
belong to~$\Fix\cup C$.
Let~$\gamma(C)$ be $\set{x^n_C~:~n\in\ints,~x_C^n\not\in\Fix}$.
According to the previous remark~$\gamma(C)\subseteq C$.

We now define~$\gamma$ to be the union of~$\gamma(C)$ for~$C$ ranging
over equivalence classes of~$\sim$. Let us prove that this~$\gamma$
satisfies the conclusion of the lemma.

Let~$x<y$ be in~$\alpha$ such that~$]x,y[\cap\gamma$ contains two
distinct elements.
If~$]x,y[$ contains two elements $x'<y'$ nonequivalent for~$\sim$,
there is a fixpoint in~$[x',y']\subseteq]x,y[$.
It follows by Fact~\ref{fact:proplr} that~$R(x,y)$.
Else~$]x,y[$ is included in some equivalence class~$C$ of~$\sim$.
Thus, the two elements in~$]x,y[$ are of the form~$x_C^n$ and~$x_C^m$
for~$n<m$. Since~$x_C^n<x_C^{n+1}\leq x_C^m$, $x_C^{n+1}=r(x_C^n)$ belongs to~$]x,y[$.
By Fact~\ref{fact:proplr}, $R(x,y)$.

Let~$x<y$ be in~$L$ such that $R(x,y)$.
If~$x\not\sim y$ then by definition~$\Fix\cap[x,y]$ is nonempty.
And since~$\Fix\subseteq\gamma$, $[x,y]\cap\gamma$ is nonempty.
Else~$x\sim y$. Let~$C$ be the equivalence class containing both~$x$
and~$y$. If~$x_C\in[x,y]$, then~$x_C$ witnesses the nonemptyness of
$\gamma\cap [x,y]$.
Else either~$x>x_C$ or~$y<x_C$. The two cases are symmetric. Let us treat
the case~$x>x_C$. By Fact~\ref{fact:proplr}, $r^\omega(x_C)\in\Fix$,
and as~$x_C\sim x$,~$x<r^\omega(x_C)$.
Hence, there exists some~$n$ in~$\nats$ such that~$x_C^n=r^n(X_C)\geq x$.
Let~$n$ be the least such natural. We have~$x^{n-1}_C<x$, and by monotonicity (Fact~\ref{fact:proplr})
$x^n_C\leq r(x)$. Overall~$x_C^n\in[x,r(x)]$.
Furthermore by Fact~\ref{fact:proplr}, $r(x)\leq y$.
This witnesses $x_C^n\in \gamma\cap[x,y]$.\qed
\end{proof}

We will also require the following lemma\footnote{In fact, the weaker result needed is the existence of a
mapping~$c:\alpha\rightarrow\{0,\dots,k-1\}$ such that for all~$x<y$ in~$\alpha$
with~$c(x)=c(y)=0$, $c([x,y])=\{0,\dots,k-1\}$. It happens to be much easier to establish than
Lemma~\ref{lemma:modulo}.}.
\begin{lemma}\label{lemma:modulo}
For every linear ordering~$\alpha$ and every natural~$k$,
there exists a mapping~$c:\alpha\rightarrow\{0,\dots,k-1\}$
such that for every~$x<y$ in~$\alpha$ with~$c(x)=c(y)$, $c([x,y])=\{0,\dots,k-1\}$.
\end{lemma}
\begin{proof}
Let~$[k]$ denote~$\{0,\dots,k-1\}$
We first show the result for a dense linear ordering~$\beta$.
Consider the set~$M$ of partial mappings $c$ from~$\beta$ to~$[k]$
such that for every~$x<y$ with~$c(x)=c(y)$ defined, either
$c$ is injective when restricted to~$[x,y[$,
or~$c([x,y])=[k]$.
Those mappings are ordered by $c\subseteq c'$ if the domain of~$c'$ contains
the domain of~$c$, and~$c$ coincides with~$c'$ over its domain.
Consider now a chain~$(c_i)_{i\in I}$ of elements in~$M$.
It has an upper bound~$b$
defined by~$b(x) = c_i(x)$ if there is some~$i$ such that~$c_i(x)$ is defined,
else $b(x)$ is undefined. It is easy to check that~$b$ belongs also to~$M$.
By Zorn's lemma, there exists a maximal element~$m$ in~$M$.
Assume~$m$ is not defined in say, $x$. Let~$Y$ be the set of
elements~$y$ such that $m$ is not defined over~$[\min(x,y),max(x,y)]$.
By definition, $x\in Y$.
There are four cases depending on whether~$x$ is the minimal (resp. the maximal)
element of~$Y$.
If~$x$ is neither the minimal nor the maximal element, this means there
exists $y<x<z$ in~$Y$. By density, we can construct a $\zeta$-indexed growing sequence
$(x_i)_{i\in\ints}$ included in~$Y$. Define then~$m'$ to coincide everywhere with~$m$,
but over the~$x_i$'s, where~$m'(x_i)$ is set to be the remainder of~$i$ modulo~$k$.
By construction~$m'$ belongs to~$M$, contradicting the maximality of~$m$.
If~$Y$ is~$[x]$, set~$m'$ to coincide everywhere with~$m$ but for~$x$, where $m'(x)=0$.
Once more, $m'$ belongs to~$M$, this time by remarking that every value in~$[k]$
is mapped by~$m$ infinitely close to the left and to the right of~$x$.
This contradicts the maximality of~$m$.
The other possibilities for~$Y$ are just combinations of the two above.
Hence~$m$ has to be defined everywhere, which means by density of~$\beta$
that the conclusion of the lemma holds for every dense linear ordering.

At this point, the easiest way to conclude the proof is to prove for every~$n$ in~$[k]$
and every scattered nonempty linear ordering~$\beta$, that
there exists a mapping~$c_{\beta,n}$ satisfying the conclusion of the lemma, such that
$c_{\beta,n}^{-1}(n)$ is nonempty. This can be easily done with the help of
Hausdorff's theorem (see e.g. chapter 5 in \cite{Rosenstein82}).
Then, one uses the fact that every linear ordering~$\alpha$
is a dense sum of scattered linear orderings (Theorem~4.9 in \cite{Rosenstein82}), i.e:
\begin{align*}
\alpha=\sum_{x\in\gamma}\beta_x\quad\text{with~}\gamma~\text{dense,~and~all the~}\beta_x~\text{are scattered and pairwise disjoint}.
\end{align*}
Then, using the case of a dense linear ordering above, we have a mapping~$d$
from~$\gamma$ to~$[k]$ satisfying the conclusion of the lemma.
Define now~$c$ over~$\alpha$ by~$c(x)=c_{\beta,d(\beta)}(x)$
for~$\beta\in\gamma$ with~$x\in\beta$. This mapping~$c$ fulfills the conclusion of
the lemma.
\qed
\end{proof}

\subsection{Proof of the statement}
\label{subsection:simon-general-proof}

We assume here the reader used to standard semigroup theory,
and in particular Green's relations.
The reader can refer to
\cite{lallement79,pinvariete84,pinvariety86} for a presentation of
the subject. Some definitions and facts are presented below.

Below, $\sigma$ denotes the additive labelling from the complete linear ordering $\alpha$
to the finite semigroup~$(S,.)$ of Theorem~\ref{theorem:general-simon}.
We denote by~$\beta$ a subordering of~$\alpha$.
We slightly abuse the notation, and write $(\beta,\sigma)$ for~$(\beta,\sigma|_\beta)$
in which~$\sigma|_\beta$ is the additive labelling obtained by restricting~$\sigma$ to~$\beta$.
We also denote by~$\sigma(\beta)$ the set $\set{\sigma(x,y)~:~x<y,~x,y\in \beta}$.

\subsection*{Facts about finite semigroups and Green's relations}

We recall some definitions here, and gather some standard facts concerning finite semigroups.

Given a semigroup~$S$, $S^1$ denotes the monoid~$S$
itself if~$S$ is a monoid, or the monoid~$S$ augmented with
a new neutral element~$1$ otherwise, thus making $S$ a monoid.

The Green's relation are defined by:
\begin{align*}
a\leq_{\gL} b & 	\quad\text{if}\quad a=cb~\text{for some~}c\text{~in~}S^1
	& a\gL b&	\quad\text{if}\quad a\leq_{\gL} b\text{~and~}b\leq_{\gL} a\\
a\leq_{\gR} b & 	\quad\text{if}\quad a=bc~\text{for some~}c\text{~in~}S^1
	& a\gR b&	\quad\text{if}\quad a\leq_{\gR} b\text{~and~}b\leq_{\gR} a\\
a\leq_{\gJ} b &		\quad\text{if}\quad a=cbc'~\text{for some~}c,c'\text{~in~}S^1
	& a\gJ b&	\quad\text{if}\quad a\leq_{\gJ} y\text{~and~}b\leq_{\gJ} a\\
a\leq_{\gH} b & 	\quad\text{if}\quad a\leq_{\gL} b\text{~and~}a\leq_{\gR} b
	& a\gH b & 	\quad\text{if}\quad a\gL b\text{~and~}a\gR b
\end{align*}

\begin{fact}
Let $a,b,c$ be in~$S$. If~$a\gL b$ then $ac\gL bc$.
If~$a\gR b$ then $ca\gR cb$.
For every~$a,b$ in~$S$, $a\gL c\gR b$ for some~$c$ iff $a\gR c'\gL b$ for some~$c'$.
\end{fact}
As a consequence of the last equivalence, one defines the last of Green's relations:
\begin{align*}
a\gD b	& \quad\text{if}\quad a\gL c\gR b ~\text{for some~}c\text{~in~}S\\
	& \quad\text{if}\quad a\gR c'\gL b ~\text{for some~}c'\text{~in~}S
\end{align*}
The key result being (here the hypothesis of finiteness of $S$
is mandatory):
\begin{fact}
$\gD=\gJ$.
\end{fact}
For this reason, we refer from now on only to $\gD$
and not~$\gJ$. However, we will use the preorder~$\leq_{\gJ}$
(which is an order over the $\gD$-classes).

An elemement $a$ in~$S$ is called \intro{regular}
if~$asa=a$ for some~$s$ in~$S$. A $\gD$-class is \intro{regular}
if all its elements are regular.
\begin{fact}\label{fact:D-regular}
A $\gD$-class $D$ is regular, iff it contains an idempotent, iff every $\gL$-class in $D$
contains an idempotent, iff every~$\gR$-class in $D$ contains an idempotent, iff there
exists $a,b$ in~$D$ such that~$ab\in D$.
\end{fact}
\begin{fact}\label{fact:LR-prolong}
For every~$a,b$ in~$D$ such that~$ab\in D$, $a\gR ab$
and~$b\gL ab$. Furthermore, there is an idempotent~$e$ in~$D$
such that~$a\gL e$ and $b\gR e$.
\end{fact}
\begin{fact}[from Green's lemma]\label{fact:size-D-class}
All~$\gH$-classes in a~$\gD$-class have the same cardinality.
\end{fact}
\begin{fact}\label{fact:H-group}
Let~$H$ be an~$\gH$-class in~$S$.
Either for all~$a,b$ in~$H$, $ab\not\in H$;
or for all~$a,b$ in~$H$, $ab\in H$,
and furthermore~$(H,.)$ is a group.
\end{fact}

\subsubsection*{Case of a group $\gH$-class.}

\begin{lemma}\label{lemma:simon-general-H}
Let~$H$ be an~$\gH$-class in~$S$ such that~$(H,.)$ is a group,
and~$\beta$ be such that~$\sigma(\beta)\subseteq H$.
Then there exists a ramseyan split  of height at most~$|H|$ of~$(\beta,\sigma)$.
\end{lemma}
\begin{proof}
Since~$(H,.)$ is a group,
it is natural to extend the definition of~$\sigma$
over~$\beta$ in the following way.
For every~$x$, let~$\sigma(x,x)$ be~$1_H$,
the neutral element of the group~$(H,.)$;
for every~$y<x$ in~$\beta$, let $\sigma(x,y)$ be~$\sigma(y,x)^{-1}$,
the inverse of~$\sigma(x,y)$ in~$H$.
As expected, this extended version of $\sigma$
satisfies for every~$x,y,z$ in~$\beta$,
$\sigma(x,z)=\sigma(x,y)\sigma(y,z)$.
Let~$n$ be a mapping  numbering the elements of~$H$
from~$1$ to~$|H|$. Fix an element~$x_0$ in~$\beta$.
Let~$s$ be defined for all~$x$ by~$s(x)=n(\sigma(x_0,x))$.

Let us show that~$s$ defined this way is indeed a ramseyan split for~$\sigma$.
Let~$x<y$ be such that~$s(x)=s(y)$, then~$\sigma(x_0,x)=\sigma(x_0,y)$
since~$n$ is a bijection from~$H$ onto~$[1,|H|]$.
Hence~$\sigma(x,y)=\sigma(x,x_0)\sigma(x_0,y)=\sigma(x_0,x)^{-1}\sigma(x_0,y)=1_H$.
Hence, given $x<y$ and~$x'<y'$ pairwise $k$-neighbours, then
$\sigma(x,y)=1_H=\sigma(x',y')=1_H^2$. \qed
\end{proof}

\subsubsection*{Case of a regular $\gD$-class.}
\begin{lemma}\label{lemma:simon-general-D}
Let~$D$ be a regular $\gD$-class in~$S$, and~$\beta$ be such that~$\sigma(\beta)\subseteq D$.
Then there exists a ramseyan split  of height at most~$|D|$ of~$(\beta,\sigma)$.
\end{lemma}
\begin{proof}
For every~$x\in\beta$ nonmaximal,
set~$r(x)$ to be the~$\gR$-class of~$\sigma(x,z)$ for some~$z>x$;
this value is independant of the choice of~$z$ according to Fact~\ref{fact:LR-prolong}.
Similarly, for every~$x$ in~$\beta$ nonminimal,
set~$l(x)$ to be the~$\gL$-class of~$\sigma(y,x)$ for some~$y<x$.
If~$\beta$ has a maximal element~$M$,
choose~$r(M)$ to be such that~$l(M)\cap r(M)$ is a subgroup of~$S$;
this is possible according to Fact~\ref{fact:D-regular}.
Similarly if~$\beta$ has a minimal element~$m$,
choose~$l(m)$ such that~$l(m)\cap r(m)$ is a subgroup of~$S$.
Set for all~$x$ in~$\beta$, $h(x)=l(x)\cap r(x)$.

We claim that for every~$x$ in~$\beta$, $h(x)$ is a subgroup of~$S$.
Indeed, if~$x$ is either the minimal or the maximal element of~$\beta$, this follows from the definition
of~$r(M)$ and~$l(m)$.
Else, there exists $y,z$ such that $y<x<z$. Let~$a$ be~$\sigma(y,x)\in l(x)$
and~$b$ be~$\sigma(x,z)\in r(x)$.
By Fact~\ref{fact:LR-prolong}, since $ab=\sigma(y,z)\in D$,
there exists an idempotent~$e$ in~$D$ such that~$a\gL e$ and $b\gR e$;
i.e. $e\in h(x)$. And by Fact~\ref{fact:H-group}, $h(x)$ is a subgroup of~$S$.
The claim holds.

According to Fact~\ref{fact:size-D-class},
there is a natural number~$N$ such that all~$\gH$-classes included
in~$D$ have cardinal~$N$.
Let~$H_1,\dots,H_d$ be the $\gH$-classes included in~$D$
which are subgroups of~$S$.
For~$k$ in~$\{1,\dots,d\}$, set~$\beta_k$
to be~$\set{x\in\beta~:~h(x)=H_k}$.
By fact~\ref{fact:LR-prolong}, $\sigma(\beta_k)\subseteq H_k$.
By Lemma~\ref{lemma:simon-general-H},
there exists a ramseyan  split~$s_k$
for~$(\beta_k,\sigma)$ of height at most~$|H_k|=N$.

We set now for all~$x$ in~$\beta$,
$s(x)$ to be~$kN+s_k(x)$ where~$k$ is such that~$x\in\beta_k$.
Let us establish that~$s$ is a ramseyan split for~$(\beta,\sigma)$.
Let~$x<y$ and~$x'<y'$ be such that~$s(x)=s(y)=s(x')=s(y')$.
By definition of~$s$, $x,y,x',y'$ belong to the same~$\beta_k$.
Furthermore, since $s(x)=s(y)=s(x')=s(y')$,
we have~$s_k(x)=s_k(y)=s_k(x')=s_k(y')$.
Hence, by ramseyanity of~$s_k$ over~$(\beta_k,\sigma)$,
$\sigma(x,y)=\sigma(x',y')=\sigma(x,y)^2$.
We conclude that the mapping~$s$ is a ramseyan
split for~$(\beta,\sigma)$.
Its height is bounded by~$dN\leq|D|$.
\qed
\end{proof}

\subsubsection*{The general case for ordinals: proof of Theorem~\ref{theorem:simon-split}.}~

For this last part of the proof, one has to provide factorisations on ordinals
where the minimal value has ben removed. Without this, one does not obtain
the bound of~$|S|$ announced.
Hence, given a linear well-ordering~$\beta$, one denotes by~$\dot\beta$
the linear ordering $\beta\setminus\{0_\beta\}$.

\begin{lemma}\label{lemma:simon-general-E}
Let~$E\subseteq S$ be a $\gD$-closed subset of~$S$
and $\beta\subseteq\alpha$ be such that~$\sigma(\beta)\subseteq E$.
Then there exists a ramseyan split  of height at most~$|E|$
of~$(\dot\beta,\sigma)$.
\end{lemma}

\begin{proof}
The proof is done by induction on the size of~$E$.
If~$E$ is empty, then~$\beta$ contains at most one element.
Hence~$\dot\beta$ is empty. We can give a split of height~$0$
over the empty linear ordering.

Else, let~$D$ be a minimal~$\gD$-class in~$E$
(for the~$\leq_{\gJ}$-order).
Let~$\gamma\subseteq\beta$ be the least set satisfying:
\begin{itemize}
\item $0_\beta\in \gamma$, where~$0_\beta$ is the minimal element of~$\beta$,
\item if~$x\in \gamma$ then~$\min\{y>x~:~\sigma(x,y)\in D\}\in \gamma$.
\end{itemize}
It is not difficult to check that the following fact holds.
\begin{fact}\label{fact:property-gamma-well}
For every~$x,y$ in~$\beta$,
if~$]x,y]\cap\gamma$ is empty, then~$\sigma(x,y)\not\in D$.
If~$[x,y]\cap\gamma$ contains two elements, then~$\sigma(x,y)\in D$.
\end{fact}
Define the equivalence relation~$\sim$ over~$\beta$
by~$x\sim y$, if~$]x,y]\cap\gamma=\emptyset$ for~$x<y$
and closed under reflexivity and symmetry.
Let~$\eta$ be an equivalence class for~$\sim$.
By Fact~\ref{fact:property-gamma-well}, $\sigma(\eta)\cap D=\emptyset$.
Hence, one can apply the induction hypothesis and obtain a
ramseyan split~$s_{\dot\eta}$ for~$(\dot\eta,\sigma)$
of height at most $|E|-|D|$.
Remark that~$\dot\eta=\eta\setminus\gamma$.

At this point, two cases may happen depending on the regularity of~$D$.
If~$D$ is not regular, then~$\gamma$ contains at most~$2$ elements,
Indeed, assume~$x<y<z$ in~$\gamma$, then~$\sigma(x,y),\sigma(y,z)$
and~$\sigma(x,y)\sigma(y,z)=\sigma(x,z)$ belong to~$D$.
By Fact~\ref{fact:D-regular}, $D$ would be regular. A contradiction.
Define~$s_{\dot\beta}$ over~$\dot\beta$ by
$s(x)=1$ for~$x\in\gamma$, else $s(x)=s_{\dot\eta}(x)+1$
for~$\eta$ the equivalence class of~$x$.
This split is ramseyan since the value~$1$ is used
at most once (in $\dot\gamma$), and the ramseyanity is inherited from
the induction hypothesis elsewhere. By induction hypothesis,
this split has height at most~$|E|-|D|+1\leq|E|$.

Finally, if~$D$ is regular. We have
$\sigma(\gamma)\subseteq D$. By Lemma~\ref{lemma:simon-general-D}
we obtain ramseyan split~$s_\gamma$ of height at most~$|D|$
for~$(\gamma,\sigma)$. Then define~$s$ over~$\dot\beta$ by
$s(x)=s_\gamma(x)$ for~$x\in\gamma$, else $s(x)=|D|+s_\eta(x)$
for~$\eta$ the equivalence class of~$x$.
It follows from the definition that~$s$ is a
ramseyan split of~$(\dot\beta,\sigma)$ of height at most~$|E|-|D|+|D|=|E|$.
\qed
\end{proof}

We can now conclude the proof of Theorem~\ref{theorem:simon-split}.
\begin{proof}
Given an ordinal~$\alpha$, and an additive labelling~$\sigma$
from~$\alpha$ to~$S$.
Fix a value~$a_0$ in~$S$, construct
the linear ordering~$\alpha'=1+\alpha$, where $1$ is a linear ordering
containing the single element $0$.
Set~$\sigma'(x,y)$ for~$x<y$ in $\alpha$ to be~$\sigma(x,y)$,
and set~$\sigma'(0,y)$ to be~$a_0.\sigma(0_\alpha,y)$.
Defined like this, $\sigma'$ is an additive labelling from~$\alpha'$
to~$S$. By Lemma~\ref{lemma:simon-general-E},
there exists a ramseyan split~$s$ for~$(\dot\alpha',\sigma')$
of height at most~$|S|$.
By construction of~$\alpha'$ and~$\sigma'$, $s$ is
also  a ramseyan split for~$(\alpha,\sigma)$.
\qed
\end{proof}

\subsubsection*{The general case for complete orderings: proof of Theorem~\ref{theorem:general-simon}.}~

Theorem~\ref{theorem:general-simon} follows directly
from the following lemma, with~$E=S$.
\begin{lemma}\label{lemma:simon-general-E-complete}
Let~$E\subseteq S$ be a $\gD$-closed subset of~$S$
and $\beta\subseteq\alpha$ be complete and such that~$\sigma(\beta)\subseteq E$.
Then there exists a ramseyan split  of height at most~$3|E|$
of~$(\beta,\sigma)$.
\end{lemma}
\begin{proof}
We assume wlog that~$\beta$ is nonempty.
The proof is done by induction on the size of~$E$.
Let~$D$ be a minimal~$\gD$-class in~$E$
(for the~$\leq_{\gJ}$-order).
We define a binary relation~$R$ over~$\beta$ by:
for every~$x<y$ in~$\beta$, $R(x,y)$ if~$\sigma(x,y)\in D$.
Since~$D$ is a minimal~$\gD$-class, this relation is upward closed;
we can apply Lemma~\ref{lemma:firstS} and
obtain a set~$\gamma$ satisfying its conclusion.

Define the equivalence relation~$\sim$ over~$\beta\setminus\gamma$
by~$x\sim y$, if~$[x,y]\cap\gamma=\emptyset$ for~$x<y$
and closed under reflexivity and symmetry.
Let~$\eta$ be an equivalence class for~$\sim$.
By Lemma~\ref{lemma:firstS} from which is obtained~$\gamma$,
$\sigma(\eta)\cap D=\emptyset$.
Hence, one can apply the induction hypothesis and obtain a
ramseyan split~$s_\eta$ for~$(\eta,\sigma)$.
At this point, two cases may happen depending on the regularity of~$D$.

If~$D$ is not regular, then~$\gamma$ contains at most~$2$ elements
(same argument as in the case of~$\alpha$ being an ordinal.
Let us treat the case of~$\gamma$ containing two elements $x_0< x_1$
(the case of~$\gamma$ being empty or a singleton can be deduced from it).
The equivalence~$\sim$ has at most three equivalence classes,
$\eta=(-\infty,x_0[$, $\eta'=]x_0,x_1[$, and~$\eta''=]x_1,+\infty)$.
We can apply the induction hypothesis with~$\sigma(\eta)\subseteq E\setminus D$
(resp. $\sigma(\eta')\subseteq E\setminus D$
and~$\sigma(\eta'')\subseteq E\setminus D$)
and obtain a ramseyan split $s_\eta$ for~$(\eta,\sigma)$
(resp. $s_{\eta'}$ for~$(\eta',\sigma)$ and~$s_{\eta''}$ for~$(\eta'',\sigma)$)
of height at most~$3(|E|-|D|)$.
We construct~$s$ over~$\beta$ by~$s(x)=s_\eta(x)+2$ if~$x\in\eta$,
$s(x_0)=1$, $s(x)=s_{\eta'}(x)+2$
if~$x\in\eta'$, $s(x_1)=2$, and~$s(x)=s_{\eta''}(x)+2$ for~$x\in\eta''$.
It follows from the definition that~$s$ is a
ramseyan split of~$(\beta,\sigma)$ of height at most~$3(|E|-|D|)+2\leq 3|E|$.

Else, if~$D$ is regular,
we apply Lemma~\ref{lemma:modulo} on~$\gamma$ with~$k=3$
and obtain a mapping~$c:\gamma\rightarrow\{0,1,2\}$
satisfying the conclusions of Lemma~\ref{lemma:modulo}.
By Lemma~\ref{lemma:firstS}, $\sigma(c^{-1}(0))\subseteq D$.
We can apply Lemma~\ref{lemma:simon-general-D} to~$c^{-1}(0)$,
obtaining a ramseyan split~$s'$ for~$(c^{-1}(0),\sigma)$
of height at most~$|D|$.
Let~$x$ be in~$\beta$, we define
\begin{align*}
s(x)&=	\begin{cases}
	s'(x)\quad&\text{if}~x\in \gamma,~\text{and}~c(x)=0\\
	|D|+c(x)~&\text{if}~x\in \gamma,~c(x)\in\{1,2\}\\
	s_\eta(x)+|D|+2\quad&\text{if}~x\not\in\gamma,~\text{and}~\eta~\text{is the $\sim$-equivalence class of}~x.
	\end{cases}
\end{align*}

Let us first remark that the values corresponding to the first case of the definition
range in~$[1,|D|]$ (def. of~$s'$). The values of the second case lie in~$[|D|+1,|D|+2]$ by construction.
Finally, the
values provided by the last case lie all
in $[|D|+3,|D|+2+3(|E|-|D|)]$,
which is included in~$[|D|+3,3|E|]$.

We have to prove the ramseyanity of~$s$.
Let~$x<y$ and~$x'<y'$ be pairwise $k$-neighbours for some~$k$.
If~$k\in[1,|D|]$, we are in the first case of the definition of~$s$,
and $\sigma(x,y)=\sigma(x',y')=\sigma(x,y)^2$
by ramseyanity of~$s'$.
If~$k\in[|D|+1,|D|+2]$, then~$c(x)=c(y)$ and
by Lemma~\ref{lemma:modulo},
there is some~$z$ in~$]x,y[$
with~$c(z)=0$. This implies~$s(z)\leq|D|$, contradicting
the `$k$-neighbourity' of~$x$ and~$y$.
Finally if~$k\geq |D|+3$, since~$x,y,x'$ and~$y'$
are~$k$-neighbours, they all lie in the same $\sim$-equivalence class~$\eta$.
And $\sigma(x,y)=\sigma(x',y')=\sigma(x,y)^2$ by
 ramseyanity of~$s_\eta$.
\qed
\end{proof}

\section{Application to countable scattered linear orderings}
\label{section:scattered}

In this section, we use Theorem~\ref{theorem:general-simon}
for giving a new simplified proof of Theorem~\ref{theorem:diamond-equivalence}
(known from~\cite{cartonrispal2005}).
We first briefly recall some facts about scattered
linear orderings in Section~\ref{subsection:recall-scattered}
and define the corresponding notions for words.
Then we introduce automata on countable scattered words in Section~\ref{subsection:automata-scattered}
and the corresponding algebraic definition
of a $\diamond$-semigroup
in Section~\ref{subsection:diamond-semigroups}.
In Section~\ref{subsection:equivalence-scattered},
we prove Theorem~\ref{theorem:diamond-equivalence}.

This section is independant from the subsequent ones.

\subsection{Scattered linear orderings}
\label{subsection:recall-scattered}

A linear ordering $\alpha$ is \intro{dense}
if for every~$x<y$ in~$\alpha$, there exists~$z$ in~$]x,y[$.
A linear ordering is \intro{scattered}
if it is not dense on any subordering.
For instance~$(\rationals,<)$ and~$(\reals,<)$ are dense,
while~$(\nats,<)$ and~$(\ints,<)$ are scattered.
Being scattered is preserved under taking a subordering.
A scattered sum of scattered linear orderings also yields
a scattered linear ordering.
Every ordinal is scattered.
Furthermore, if~$\alpha$ is scattered,
then~$\cuts\alpha$ is scattered. And if~$\alpha$ is countable
and scattered, then~$\cuts\alpha$ is also countable and scattered.

Given an alphabet~$A$, we denote by~$A^\diamond$
the set of words indexed by a countable scattered
linear ordering.
Given a language~$L\subseteq A^\diamond$,
$L^\omega$ represents the set of words
of the form~$\prod\set{u_i~:~i\in\omega}$
where all the~$u_i$'s belong to~$L$. One defines
similarly~$L^{-\omega}$ and~$L^\zeta$.

A standard way for proving results on scattered linear
orderings is to use the theorem of Hausdorff
(chapter~5 of~\cite{Rosenstein82} is dedicated to the subject).
It establishes a general way of decomposing
scattered linear orderings. Hausdorff's theorem is
a key tool in the original proof of
Theorem~\ref{theorem:diamond-equivalence} \cite{cartonrispal2005}.
We avoid it below; instead,
we use the following lemma which provides a kind
of induction principle for scattered linear orderings.
It essentially says that an equivalence relation such
that any two sets of equivalent elements are contiguous
(there is nothing in between) are equivalent,
then the relation contains is trivial.
\begin{lemma}\label{lemma:scattered-induction}
Given a scattered linear ordering $\alpha$
and an equivalence relation~$R$ over~$\alpha$
satisfying:
\begin{align*}
\text{for all}~X<Y,~\text{with}~X^2\subseteq R,~\text{and}~Y^2\subseteq R,
\qquad \bigcap\limits_{x\in X,~y\in Y}]x,y[=\emptyset\quad\text{implies}
~(X\cup Y)^2\subseteq R~;
\end{align*}
Then $R=\alpha^2$.
\end{lemma}
\begin{proof}
Consider the set~$S$ of equivalence relations included in~$R$
such that every equivalence class is convex.
It is nonempty since the equality relation over~$\alpha$
belongs to~$S$. Order~$S$ by inclusion.
Given a chain in~$S$, the union of all relations in the
chain is itself an element of~$S$: the chain has an upper bound in~$S$.
Then, according to Zorn's lemma, there is a maximal element~$\sim$ in~$S$.
Since~$\alpha$ is scattered and~$\sim\in S$, $\alpha/_\sim$ is itself a scattered linear
ordering. Assume that it has two distinct equivalence classes. Since~$\alpha/_\sim$
is scattered, there are two equivalence classes~$X$ and~$Y$
--- choose wlog~$X<Y$ --- such that there is no other equivalence class~$Z$
with~$X<Z<Y$. This follows that~$\cap_{x\in X,~y\in Y}]x,y[=\emptyset$.
Applying the hypothesis leads to $(X\cup Y)^2\subseteq R$, and consequently
$(\sim\cup (X\cup Y)^2)\in S$.
It contradicts
the maximality of~$\sim$.\qed
\end{proof}

\subsection{Automata over countable scattered linear orderings}
\label{subsection:automata-scattered}

In this section, we define priority automata and show how
they accept words indexed by countable scattered linear orderings.
Those automaton were introduced in~\cite{bruyerecarton2001},
but in their `Muller' form, while here we adopt the `parity-like' approach.
\begin{definition}
A \intro{priority automaton} $\mathcal{A}=(Q,A,I,F,p,\delta)$
consists of a finite set of \intro{states} $Q$,
a finite alphabet~$A$, a set of \intro{initial states}~$I$,
a set of final states~$F$, a \intro{priority mapping}~$p: Q\mapsto[1,N]$ ($N$ being a natural)
and a \intro{transition relation}
$\delta\subseteq (Q\times A\times Q)\uplus([1,N]\times Q)\uplus(Q\times[1,N])$.
\end{definition}

A \intro{run} of the automaton~$\aut$ over an $\alpha$-word~$u$
is a mapping~$\rho$ from~$\cuts\alpha$ to~$Q$ such that for all cuts~$c,c'$:
\begin{itemize}
\item if~$c'$ is the successor of~$c$ through~$x$, then
	$(\rho(c),u(x),\rho(c'))\in\delta$,

\item if~$c$ is a left limit, then $(k,\rho(c))\in\delta$
where~$k=\max\bigcap\limits_{c'<c} p(\rho(]c',c[))$,
\item if~$c$ is a right limit, then $(\rho(c),k)\in\delta$ where~$k=\max\bigcap\limits_{c'>c} p(\rho(]c,c'[))$.
\end{itemize}
The first case corresponds to standard automata on finite words: a transition links one state
to another while reading a single letter in the word.
The second case verifies that the highest priority
appearing infinitely close to the left of~$c$
corresponds to a transition. The third case is symmetric.
An $\alpha$-word~$u$ is \intro{accepted} by~$\aut$ if there is a run~$\rho$ of~$\aut$
over~$u$ such that~$\rho(\bot)\in I$ and~$\rho(\top)\in F$.

\begin{example}
Consider the automaton with states~$\set{q,r}$,
alphabet~$\set a$,
initial states~$\set{q,r}$, final state~$q$,
priority mapping constant equal to~$0$
and transitions $\set{(q,a,q),(q,a,r),(0,q),(r,0)})$.
It accepts those words in~$\set a^\diamond$ which have a complete domain.
For this, note that a linear ordering is complete iff no cut
is simultaneously a left and a right limit.

Consider a word~$u\in\set a^\diamond$ which has a
complete domain~$\alpha$. For~$c\in\cuts\alpha$,
set~$\rho(c)$ to be~$q$ if~$c$ is~$\top$ or if~$c$ has a successor,
else $\rho(c)$ is~$r$. Under the hypothesis of completeness, it is
simple to verify that~$\rho$ is a run witnessing the acceptance of the word.
Conversely, assume that there is a run~$\rho$ over the $\alpha$-word~$u$
with~$\alpha$ not complete.
There is a cut~$c\in\cuts\alpha$ which is both a left and a right limit.
If~$\rho(c)$ is~$r$, then, as~$c$ is a left limit, there is no corresponding transition;
else if~$\rho(c)$ is~$q$ the same argument apply to the right of~$c$. In both cases
there is a contradiction.
\end{example}

The languages accepted by priority automata are closed under union, intersection, concatenation,
projection and exponentiation by~$\omega$ and $-\omega$ \cite{bruyerecarton2001}.
They also admit an equivalent form of regular expressions \cite{bruyerecarton2001}
and their emptyness problem is decidable.
A consequence of Theorem~\ref{theorem:diamond-equivalence} below
is their closure under complementation (originally proved in~\cite{cartonrispal2005},
in \cite{cartonrispal05finiterank} for a particular case).

\subsection{On $\diamond$-semigroups}
\label{subsection:diamond-semigroups}

Finite semigroups are known to have the same `expressive power'
as finite state automata.
This approach has been extended to languages of $\omega$-words
while introducing $\omega$-semigroups in~\cite{perrinpin95}.
Then Bedon and Carton generalized it to words indexed by countable
ordinals in~\cite{bedoncarton98}, the corresponding
algebraic object being called an $\omega_1$-semigroup.
Finally, Carton and Rispal have introduced $\diamond$-semigroups
for describing languages of words indexed by scattered linear orderings.

Formally, a $\diamond$-semigroup $(s,\pi)$ is a set equipped with an operator~$\pi$
mapping~$S^\diamond$ to~$S$ which satisfies:
\begin{itemize}
\item for all~$s\in S$, $\pi(s)=s$, and,
\item for all countable scattered linear ordering~$\alpha$
	and families~$(u_i)_{i\in\alpha}$ of words in~$S^\diamond$,
	$$\textstyle
	\pi(\prod\set{\pi(u_i)~:~i\in\alpha})=\pi(\prod\set{u_i~:~i\in\alpha})\ .
	$$
\end{itemize}
Those properties express the fact that $\pi$ is a generalized product operator: more precisely, the rules
correspond to a generalized form of associativity. For instance,
for every~$u,v,w$ in~$S$, $\pi(u\pi(vw))=\pi(uvw)=\pi(\pi(uv)w)$.
In this sense, every $\diamond$-semigroup can be seen as a semigroup
with the product defined by~$u.v=\pi(uv)$.
The free $\diamond$-semigroup generated by a finite alphabet~$A$
is~$(A^\diamond,\prod)$.

Given two $\diamond$-semigroups~$(S,\pi)$ and~$(S',\pi')$,
a mapping~$\varphi$ from~$S$ to~$S'$ is
a \intro{morphism of $\diamond$-semigroups} if for every scattered
linear ordering~$\alpha$, and every~$(x_i)_{i\in\alpha}$ in~$S$,
$\varphi(\pi(\prod\set{x_l~:~l\in\alpha})) = \pi'(\prod\set{\varphi(x_l)~:~l\in\alpha})$.
A language~$K\subseteq A^\diamond$ is \intro{$\diamond$-recognizable}
if there exists a morphism of $\diamond$-semigroups from~$A^\diamond$
to a finite $\diamond$-semigroup saturating~$K$;
i.e. such that $\varphi^{-1}(\varphi(K))=K$.
As usual with recognizability, $\diamond$-recognizable languages
are closed under union, intersection and complementation.

From now, we denote~$\pi(uv)$ simply by~$uv$. More generally,
given a word~$u$ in~$S^\diamond$, we do not distinguish between~$u$
and~$\pi(u)$. Similarly, we abbreviate
$\pi(\prod\set{u~:~i\in(\nats,<)})$ by~$u^{\omega}$
and~$\pi(\prod\set{u~:~i\in(-\nats,<)})$ by~$u^{-\omega}$.
We also denote by~$u^\zeta$ the value~$u^{-\omega}u^\omega$.

\begin{example}
Consider the set~$S=(\set{0,1}\times\set{0,1})\uplus\set{\bot}$. Define the product~$.$
and the exponent mappings~$\omega$ and~$-\omega$ by,
for every~$x$ in~$S$ and~$a,b,a',b'$ in $\set{0,1}$,
\begin{align*}
\bot x&=x\bot=\bot & (a,b)(a',b')&=	\begin{cases}
				\bot&\ \text{if}~b=a'=1\\
				(a,b')&\ \text{else}
				\end{cases}\\
\bot^\omega&=(1,1)^\omega=\bot
	&(a,b)^\omega&=	\begin{cases}
			\bot&\text{if}~a=b=1\\
			(a,1)&\ \text{else}
			\end{cases}\\
\bot^{-\omega}&=(1,1)^{-\omega}=\bot
	&(a,b)^{-\omega}&=\begin{cases}
			\bot&\text{if}~a=b=1\\
			(1,b)&\ \text{else.}
			\end{cases}
\end{align*}
Using Theorem~10 in~\cite{cartonrispal2005}, this
$(S,.)$ together with the mappings~$\omega$ and~$-\omega$
defines uniquely a $\diamond$-semigroup~$(S,\pi)$.

Let~$u$ be in $\set{a}^\diamond$ of domain~$\alpha$.
Set~$\varphi(u)$ to be~$\bot$ if~$\alpha$ is not complete.
If~$\alpha$ is complete, set~$\varphi(u)$ to be~$(a,b)$ where~$a=0$
if~$\alpha$ has a minimal element, else~$a=1$,
and~$b=0$ if~$\alpha$ has a maximal element,
else~$b=1$. This~$\varphi$
is a morphism from~$(\set{a}^\diamond,\prod)$
to~$(S,\pi)$.
It follows that the set of words in~$\set{a}^\diamond$
of complete domain is $\diamond$-recognizable:
it is equal to~$\varphi^{-1}(\set{0,1}\times\set{0,1})$.
\end{example}

\subsection{Equivalence of representations}
\label{subsection:equivalence-scattered}

The following theorem was proved in~\cite{cartonrispal2005}\footnote{In fact,
the present theorem differs in the use of priority automata
in place of automata using Muller condition in limit transitions.
For this reason the result here is new; but for a nonessential reason.}.
A direct consequence of it is the closure under complementation of
the languages of words indexed by scattered linear orderings
accepted by priority automata.
\begin{theorem}[\cite{cartonrispal2005}]\label{theorem:diamond-equivalence}
Let~$A$ be a finite alphabet.
A language~$L\subseteq A^\diamond$
is accepted by a priority automaton if and only if it is
$\diamond$-recognizable.
\end{theorem}
The left to right implication is standard: one constructs
a $\diamond$-semigroup which captures all the possible
behaviours of the automata over a word. Then there
is no choice on the definition of the product and the morphism.

The difficult direction is, given a $\diamond$-recognizable
language, to construct a priority automaton accepting it.
The contribution here is to show that a natural way of
constructing such an automaton is to follow the
structure of a ramseyan split.
Let us fix a $\diamond$-semigroup~$(S,\pi)$ and a morphism
of~$\diamond$-semigroups~$\varphi$ from~$(A^\diamond,\prod)$ to~$(S,\pi)$.
By closure of priority automata
under union, it is sufficient to show that
for every~$c\in S$ the language $\varphi^{-1}(c)$
is acepted by a priority automaton.

Let~$k$ be a natural number, set~$L_k$
to be the set of words~$u$ such that~$\varphi_u$
admits a ramseyan  split of height at most~$k$.
We show by induction on~$k$ that for every~$c\in S$,
the language~$L_{c,k}=L_k\cap\varphi^{-1}(c)$ is accepted by an automaton.
According to Theorem~\ref{theorem:general-simon}
we have~$\varphi^{-1}(c)=L_{c,3|S|}$.
We also use the intermediate language~$\SD(e,k)$
for~$e$ an idempotent of~$S$
which is the set of words~$u$ of domain~$\alpha$
admitting a ramseyan split~$s$ of height at most~$k$, such that
$s(\bot_\alpha)=s(\top_\alpha)=1$ and~$\varphi(u)=e$
(in particular, $\SD(e,k)\subseteq L_{e,k}$).

The following lemma reduces the problem from describing the language~$L_{c,k}$
to describing languages of the form~$\SD(e,k)$.
\begin{lemma}
Let $u\in A^\diamond$ be a word of at least two letters.
Then~$u$ belongs to~$L_{c,k+1}$
iff there exists $a,b,e$ in~$S$ and~$\gamma\in\set{0,1,\omega,-\omega,\zeta}$
such that~$e^2=e$, $c=ae^\gamma b$ and
$u\in L_{a,k}(\SD(e,k+1))^\gamma L_{b,k}$
(with the convention that~$xy^0z=xz$).
\end{lemma}
\begin{proof}
From left to right.
Let~$u$ be an~$\alpha$-word in~$A^\diamond$
of length at least~$2$, and let~$s$ be a ramseyan split
of height at most~$k+1$ of $(\cuts\alpha,\varphi_u)$.
We argue on the nature of~$s^{-1}(1)$.

If~$s^{-1}(1)$ is empty, then choose arbitrarily a cut~$c$ in~$\cuts\alpha^*$,
and set a new value of~$1$ to $s(c)$. This modified~$s$ is still a ramseyan split
of height~$k+1$ of $(\cuts\alpha,\varphi_u)$. And we can apply the next case for which~$s^{-1}(1)$
is a singleton.

If~$s^{-1}(1)$ is a singleton~$\{c\}$, let~$v$ be~$u$ restricted to positions
to the left of~$c$, and~$w$ be~$u$ restricted to positions
to the right of~$c$. Obviously~$u=vw$,
and we have $u\in L_{\varphi(v),k}e^0 L_{\varphi(w),k}$
for any idempotent~$e$.

Else~$s^{-1}(1)$ contains at least two elements.
There are four cases depending on the existence of a minimal (resp. a maximal)
element in~$s^{-1}(1)$.
First case. If~$s^{-1}(1)$ has both a minimal element~$c$ and a maximal element~$c'$,
then let~$a=\varphi_u(\bot,c)$, $e=\varphi_u(c,c')$, and $b=\varphi_u(c',\top)$.
By definition of a ramseyan split, $e$ is an idempotent of~$S$; furthermore,
$\varphi(u)=aec$.
We obtain~$u\in L_{a,k}\SD(e,k+1) L_{b,k}$.
Second case.
If~$s^{-1}(1)$ has neither a minimal element nor a maximal element.
Let~$c$ be~$\inf(s^{-1}(1))$ and~$c'$ be~$\sup(s^{-1}(1))$.
Let~$a=\varphi_u(\bot,c)$, $b=\varphi_u(c',\top)$.
Using the countability of $\cuts\alpha^*$,
we have a $\zeta$-indexed sequence $\cdots<x_n<x_{n+1}<\cdots$ in~$s^{-1}(1)$,
such that~$\inf\{x_i~:~i\in\zeta\}$ is~$c$, and~$\sup\{x_i~:~i\in\zeta\}$ is~$c'$.
Let~$e$ be~$\varphi_u(x_1,x_2)$. The sequence of $x_i$'s
shows that~$\varphi_u(c,c')\in(\SD(e,k+1))^\zeta$.
Furthermore $e$ is an idempotent.
We obtain~$u\in L_{a,k}(\SD(e,k+1))^\zeta L_{b,k}$.
The two other cases are obtained as combinations of the two first one, using~$\omega$
and~$-\omega$-indexed sequences.
\qed
\end{proof}

This lemma together with the
closure properties of languages
accepted by priority automata shows that it is
sufficient to construct
an automaton accepting~$\SD(e,k+1)$.
For this, define the following languages:
\begin{align*}
M_{e,k}&=\set{u\in L_k\setminus\set\varepsilon:~\varphi(u)=e},&
M_{e,k}^{\leftarrow}&=\set{u\in L_k~:~\varphi(u)e^{-\omega}=e},\\
M_{e,k}^{\rightarrow\leftarrow}&=\set{u\in L_k~:~e^\omega\varphi(u)e^{-\omega}=e},&
M_{e.k}^{\rightarrow}&=\set{u\in L_k~:~e^\omega\varphi(u)=e}.
\end{align*}
Those languages can be obtained as unions of the~$L_{a,k}$
together with languages consisting of a single letter word, or the empty word.
Hence, by induction hypothesis there are automata accepting them.
We identify below the automaton and the language.

\begin{figure}[ht]
\begin{center}
\begin{picture}(130,27)(-50,-8)

  \node(q)(0,0){$t:n$}
  \node[Nframe=n](i)(-10,-10){$\qquad\quad$}
  \node[Nframe=n](f)(10,-10){$\qquad\quad$}

  \gasset{Nadjust=w,Nadjustdist=2,Nh=6,Nmr=1}

  \node[Nframe=n](iG)(-55,0){$\qquad\quad$}
  \node(G)(-25,0){$\quad M^{\rightarrow}_{e,k}\quad$}

  \node[Nframe=n](fD)(55,0){$\qquad\quad$}
  \node(D)(25,0){$\quad M^{\leftarrow}_{e,k}\quad$}

  \node(E)(0,15){$\quad M_{e,k}\quad$}

  \node[Nframe=n](iM)(20,15){$\qquad\quad$}
  \node[Nframe=n](fM)(80,15){$\qquad\quad$}
  \node(M)(50,15){$\quad M_{e,k}^{\rightarrow\leftarrow}\quad$}

  \drawedge[dash={1.0}0](iG,G){$n$}
  \drawedge(G,q){$\varepsilon$}
  \drawedge(q,D){$\varepsilon$}
  \drawedge[dash={1.0}0](D,fD){$n$}
  \drawqbedge(q,-25,18,E){$\varepsilon$}
  \drawqbedge(E,25,18,q){$\varepsilon$}
  \drawedge[dash={1.0}0](iM,M){$n$}
  \drawedge[dash={1.0}0](M,fM){$n$}

  \drawedge(i,q){}
  \drawedge(q,f){}
  \end{picture}
\end{center}
\caption{The automata~$\aut(e,k+1)$}
\label{figure:automata}
\end{figure}
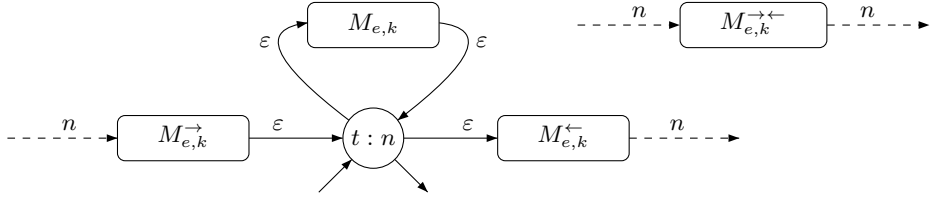

In order to accept the language~$\SD(e,k+1)$,
we construct a corresponding automaton~$\aut(e,k+1)$.
The definition of the automaton~$\aut(e,k+1)$
is depicted in Figure~\ref{figure:automata}.
This is a disjoint union of the automata
accepting~$M_{e,k},M_{e,k}^{\leftarrow},M_{e,k}^{\rightarrow}$
and~$M_{e,k}^{\rightarrow\leftarrow}$ and of a new
state~$t$ of priority~$n$; the state~$t$ being
both initial and final.
The value $n$ is chosen to be the highest priority of the automaton.
New $\varepsilon$-transitions\footnote{$\varepsilon$-transitions
		are just a commodity notation.
		And in particular there is no cycle of
		such transitions.}
are added to this construction
as depicted in Figure~\ref{figure:automata}: arrow arriving from the left
have the initial states of the automaton as destination,
while the arrows leaving to
the right have the final states of the automaton as origin.
Dashed arrows represent limit transitions. For instance the leftmost one
expresses the existence of a limit transition $(n,q)$
for~$q$ an initial state of~$M_{e,k}^\rightarrow$:
the automaton can go to state~$q$ if the maximal priority
appearing infinitely often to its left is~$n$.
The following lemma concludes the proof.

\begin{lemma}\label{lemma:main-equivalence}
	The automaton~$\aut(e,k+1)$ accepts the language~$\SD(e,k+1)$.
\end{lemma}
\begin{proof}
From right to left.  Let~$u$ be a word indexed by~$\alpha$.
Let~$s$ be a  ramseyan split of~$\varphi_u$
corresponding to the membership of~$u$
in~$\SD(e,k+1)$, i.e. such that~$s(\bot)=s(\top)=1$.

We construct a run~$\rho\in Q^{\cuts \alpha}$
in the following way ($Q$ is the set of states of
$A_{e,k+1}$).
Set~$\rho(x)=t$ whenever~$s(x)=1$.
We define $\rho$ elsewhere by copying runs of
the automata~$E_{e,k},M_{e,k}^{\leftarrow},M_{e,k}^{\rightarrow}$
and~$M_{e,k}^{\rightarrow\leftarrow}$.
More precisely,
consider a maximal interval~$I\subseteq \cuts \alpha$
such that~$s(I)\geq 2$.
Let us define~$\rho$ over~$I$.
Four cases happen depending on the nature of the interval:
$I=[x,y],[x,y[,]x,y]$
or~$]x,y[$. We treat the case of~$[x,y[$. The others are similar.

If~$I=[x,y[$, this means that~$s(x)>1$, but~$s(y)=1$.
As a consequence, there is a sequence $x_1<x_2<\dots$ in~$s^{-1}(1)$
indexed by~$\omega$ such that~$\sup\set{x_i~:~i<\omega}=x$
(this is possible because~$\cuts \alpha$ is countable).
It follows that~$\sigma(x_1,x)=e^\omega$. Furthermore (by ramseyanity),
$\sigma(x_1,y)=e$. We deduce~$e^\omega\sigma(x,y)=e$.
By induction hypothesis, we obtain that~$v$
is accepted by~$M^\rightarrow_{e,k}$.
We define~$\rho$ to replicate the corresponding
run over~$I$ using the instance of~$M^\rightarrow_{e,k}$ it contains.
We have to prove that this choice indeed produces a run.
Over~$]x,y[$ this is a correct run since the original run was
itself correct.
It remains to show the correctness of the run to the left of~$x$.
But, we already know that the
maximal priority reaching~$x$ from the left is~$n$ since
the sequence of the~$x_i$'s tends to~$x$ and by construction
correspond to a priority~$n$ which is maximal.
We conclude that there is a corresponding transition in~$A_{e,k+1}$.

From left ro right.
Let~$\rho\in Q^{\cuts \alpha}$ be a run of~$A(e,k+1)$
over~$u$ from~$t$ to~$t$.
We aim at constructing a ramseyan  split~$s$
of~$\varphi_u$
corresponding to the membership of~$u$ in~$\SD(e,k+1)$.
Let~$J$ be~$\rho^{-1}(t)$.
We set~$s(x)$ to be~$1$ over $J$.
Let~$I$ be a maximal interval which does not intersect~$J$.
Once more there are four cases: $I=[x,y],[x,y[,]x,y]$
or~$]x,y[$. We treat the case of~$[x,y[$. The others being similar.

If~$I=[x,y[$, this means that~$s(x)>1$, but~$s(x)=1$.
Let~$q$ be the state $\rho(x)$.
Since~$I$ is maximal, there exists an $\omega$-sequence
$x_1<x_2<\dots$ in~$J$ of limit~$x$. Since~$\rho(x_i)$
is~$n$ by definition, this means that the maximal
priority appearing infinitely often to the left of~$x$ is~$n$.
Hence, there must be in~$A_{e,k+1}$ a limit transition from~$n$
to~$q$. By inspecting the definition of~$A_{e,k+1}$, this means that~$q$ is either the initial state of~$M^{\rightarrow}_{e,k}$
or the initial state of~$M^{\rightarrow\leftarrow}_{e,k}$.
In~$y$, the run assumes state~$n$, but this state has been reached
by an~$\varepsilon$-transition either from the final state of~$M_{e,k}$,
or by the final state of~$M_{e,k}^{\rightarrow}$. Let~$p'$ be this state.
We know that there is a run of~$A_{e,k+1}$ from configuration~$(x,q)$ to~$(y,p')$
which does not visit state~$n$ (by definition of~$I$).
It follows that~$q$ is the initial state of~$M^\rightarrow_{e,k-1}$,
$p'$ is its final state and that the run from~$(x,q)$ to~$(y,p')$
is an accepting run of~$M_{e,k}^{\rightarrow}$.
By induction hypothesis, $\sigma_{(u|_I)}=(\varphi_u)|_I$ has factorisation
height at most~$k$. Let~$s'$ be this factorisation.
For all~$x\in I$, let $s(x)$ be~$s'(x)+1$.

Let us show that this  split is ramseyan.
Let~$x<y$ be such that~$s(x)=s(y)=k$.
For $k$-neighours with~$k\geq2$, this is inherited
from the induction hypothesis.
What remains to be shown is that for every~$x<y$
in~$J$ (i.e. $x,y$ are $1$-neighbours), $\sigma(x,y)=e$.
To make this relation reflexive and symmetric, we
consider the relation~$R$ defined by~$x R y$
if $x=y$ or $x<y$ and~$\sigma(x,y)=e$ or~$y<x$
and~$\sigma(y,x)=e$.
We want to apply Lemma~\ref{lemma:scattered-induction} on~$(J,<)$
and the relation~$R$.
Let~$X,Y\subseteq J$ be such that~$X<Y$, $X^2\subseteq R$, $Y^2\subseteq R$ and~$\cap_{x\in X,y\in Y}]x,y[\cap X=\emptyset$.
Let~$I=\cap_{x\in X,y\in Y}]x,y[$, $I$ is a maximal interval nonintersecting~$J$.

Once more there are four cases: $I=[x,y],[x,y[,]x,y]$
or~$]x,y[$. We treat the case of~$I=[x,y[$.
Fix~$x_0\in X$ and~$y_0\in Y$. We want to prove~$\sigma(x_0,y_0)=e$.
As~$x\not\in J$, there is an $\omega$-sequence $x_0<x_1<\dots$
of limit~$x$ with for all~$i$, $\sigma(x_i,x_{i+1})=e$.
It follows that~$\sigma(x_0,x)=e^\omega$.
By construction~$s$ corresponds to a run of~$M_{e,k}^{\rightarrow}$
over~$I$. It follows, by definition of~$M_{e,k}^{\rightarrow}$,
that~$e^\omega\sigma(x,y)=e$. We obtain~$\sigma(x_0,y)=e$.
Since furthermore by hypothesis, $\sigma(y,y_0)=e$,
we have~$\sigma(x_0,y_0)=e$.

Lemma~\ref{lemma:scattered-induction} concludes that for
every~$x<y$ in~$J$, $\sigma(x,y)=e$. Hence, $s$ is a ramseyan  split
for~$\varphi_u$.
\qed
\end{proof}

\section{Deterministic extension to the factorisation forest theorem}
\label{section:deterministic}

We try in this section to construct the  split from
`left to right' in a `deterministic way'. The notion of ramseyanity
is not suitable anymore in this context;
the result would be false\footnote{Consider the semigroup $(\{a,b\},.)$
defined by~$ab=aa=a$ and~$ba=bb=b$.}. It is replaced by the notion of
forward ramseyanity. The result, Theorem~\ref{theorem:simon-deterministic},
only holds for ordinals.

\subsection{The statement}

A  split~$s$ of height~$N$ is \intro{forward ramseyan}
if for every~$k=1\dots n$ and $k$-neighbours~$x<y$ and~$x'<y'$,
\begin{align*}
\sigma(x,y)&=\sigma(x,y).\sigma(x',y')\ .
\end{align*}
So in particular, $\sigma(x,y)$ is an idempotent,
but~$\sigma(x,y)$ and~$\sigma(x',y')$ may be different idempotents.
In the terminology of Green's relation,
$\sigma(x,y)$ and~$\sigma(x',y')$ are $\gL$-equivalent
idempotents.
A ramseyan  split is always forward ramseyan,
but the converse does not hold in general.

Below, we also identify the natural numbers with the
corresponding ordinal. Furthermore, for $\sigma$
an additive labelling over an ordinal~$\alpha$, and given~$\beta<\alpha$,
we denote by~$\sigma|_{\leq\beta}$ the labelling~$\sigma$
restricted to $[0,\beta]$.

\begin{theorem}\label{theorem:simon-deterministic}
Let~$(S,.)$ be a semigroup.
To every additive labelling~$\sigma$ over an
ordinal~$\alpha$,
one can associate a forward ramseyan split~$s_{\alpha,\sigma}$
of~$(\alpha,\sigma)$ of height at most~$|S|$.
Furthermore,
for every additive labellings~$\sigma$ and~$\sigma'$
over the respective ordinals~$\alpha$ and~$\alpha'$, and every ordinal
$\beta<\min\set{\alpha,\alpha'}$,
\begin{align*}
\text{if}\quad\sigma|_{\leq\beta}=\sigma'|_{\leq\beta}\quad
\text{then}~s_{\alpha,\sigma}(\beta)=s_{\alpha',\sigma'}(\beta)
\quad\quad\text{(determinism property)}\ .
\end{align*}
\end{theorem}

Furthermore, under the same hypothesis, over finite linear
orderings, the forward ramseyan  split can be computed
via monadic formul\ae.
\begin{proposition}[definable variant of Theorem~\ref{theorem:simon-deterministic}]
\label{proposition:deterministic-is-regular}
Given a finite semigroup~$(S,.)$,
there exist monadic closed formul\ae{} $\Theta_1,\dots,\Theta_{|S|}$
such that for every ordinal~$\alpha$, and additive labelling~$\sigma$
from~$\alpha$ to~$S$, the split~$s$ defined
for every~$\beta\in\alpha$ by:
\begin{align*}
s(\beta) &= n\quad\text{such that}\quad (\beta+1,\sigma|_{\leq\beta})\models\Theta_n\ ,
\end{align*}
is forward ramseyan.
\end{proposition}
\begin{proof}(Idea)
Implement the construction of the proof of Theorem~\ref{theorem:simon-deterministic}
via monadic formul\ae{}.
\qed
\end{proof}
Note, that a consequence of Proposition~\ref{proposition:deterministic-is-regular},
the mapping~$s$ satisfies the determinism property.

\subsection{Proof of Theorem~\ref{theorem:simon-deterministic}}
\label{subsection:proof-deterministic}

Once more, we perform a case analysis.

\subsubsection*{Case of a single~$\gH$-class.}

\begin{lemma}\label{lemma:simon-determinist-H}
Let~$H$ be an~$\gH$-class in~$S$ such that~$(H,.)$ is a group.
For~$\beta\subseteq\alpha$ such that~$\sigma(\beta)\subseteq H$,
there exists a ramseyan split~$s^H_{\beta,\sigma}$
of height at most~$|H|$. Furthermore $s^H$ satisfies the
determinism property.
\end{lemma}
This is exactly the proof of Lemma~\ref{lemma:simon-general-H}
in which one always chooses~$x_0$ to be~$0_\beta$.

\subsubsection*{Case of a single~$\gL$-class.}

\begin{lemma}\label{lemma:simon-determinist-L}
Let~$L$ be an~$\gL$-class in a regular $\gD$-class,
for every ordinal~$\beta$ such that~$\sigma(\beta)\subseteq L$,
there exists a ramseyan split~$s^L_{\beta,\sigma}$
of height at most~$|L|$. Furthermore $s^L$ satisfies the
determinism property.
\end{lemma}

We require the following result.
\begin{fact}\label{fact:LH}
There is an~$\gH$-class~$H\subseteq L$ which is a group,
and a mapping~$f:L\rightarrow H$ such that:
\begin{itemize}
\item for every~$a,b$ in~$L$, if~$ab\in L$ then~$f(ab)=f(a)f(b)$, and,
\item for every~$\gH$-class $H'\subseteq L$, $f|_{H'}$ is a bijection from~$H'$
	onto~$H$.
\end{itemize}
\end{fact}
\begin{proof}
Let~$H_1,\dots,H_n$ be the $\gH$-classes
included in~$L$.
By Fact~\ref{fact:H-group} we can assume that~$H_1,\dots,H_k$
are groups, while for every~$a,b$ in~$K_i$ for~$i>k$, $ab\not\in L$.
By regularity hypothesis and Fact~\ref{fact:D-regular}, $k\geq 1$.
Let~$L'=H_1\cup\dots\cup H_k$.

Let~$a,b$ be in~$L$, we claim that~$ab\in L$ iff~$b\in L'$.
Indeed, if~$b\in L'$, let~$e$ be the neutral element of the group containing~$b$.
Since~$e\gL a$, $e=xa$ for some~$x$. Hence, $b=eb=xab$, and we deduce $ab\gL b$.
Conversely, suppose~$ab$ in~$L$, then~$ab\gR a$. Hence, $a=abc$ for some~$c$.
But then $abcbc=a$. Hence $bc$ belongs to~$L'$. But~$bc\gR b$. Hence~$b\in L'$.

Let~$H$ be~$H_1$. If~$k=0$, then for all~$a,b$ in~$L$, $ab\not\in L$.
One can construct the mapping arbitrarily using
Fact~\ref{fact:size-D-class}.
Else, let~$e_i$ be the neutral element of~$H_i$ for~$i\leq k$.
Let~$i,j\leq k$. Since~$e_i\gL e_j$, $e_i=xe_j$ for some~$x$.
Hence~$e_ie_j=xe_je_j=xe_j=e_i$.
For every~$a\in L$, let~$f(a)=a e_1$.
According to the claim above, $f$ is a mapping from~$L$ to~$H_1$.
Assume $a,b$ in~$L$ such that~$ab\in L$. According to
the claim, above, $b\in L'$, i.e. $b\in H_i$ for~$i\leq k$.
Also, as~$a\gL e_i$, $a=xe_i$ for some~$x$.
We have~$f(a)f(b)=ae_1 be_1=xe_ie_1be_1=xe_ibe_1=abe_1=f(ab)$.

The fact that $f|_{H_i}$ is a bijection from~$H_i$ to~$H_1$
is known as Green's lemma.\qed
\end{proof}

We can now prove Lemma~\ref{lemma:simon-determinist-L}.
\begin{proof}
Let~$H$ and~$f$ be obtained by Fact~\ref{fact:LH}.
For~$x<y$ in~$\beta$, let~$\sigma'(x,y)$ be~$f(\sigma(x,y))$.
The first property of~$f$ makes~$\sigma'$ an additive labelling
from~$\beta$ to~$H$, such that~$\sigma(\beta)\subseteq H$.
Applying the case of a single~$\gH$-class above we obtain
a  split~$s^H_{\beta,\sigma'}$ forward ramseyan for~$(\beta,\sigma')$.
There are two different cases.

Either all the $\gH$-classes are groups.
In this case, one sets $s^L_{\beta,\sigma}$
to be~$s^H_{\beta,\sigma'}$.
Let us show that $s^L$ is forward ramseyan.
Indeed, consider~$x<y$ and~$x'<y'$
to be~$k$-neighbours for some~$k$.
This means that~$f(\sigma(x,y))$ and~$f(\sigma(x',y'))$
are equal to the neutral element~$1$ of~$H$.
Since the~$H$-class of~$\sigma(x,y)$ (\emph{resp.} of~$\sigma(x',y')$)
are groups isomorphic to~$H$, we have that~$\sigma(x,y)$
and~$\sigma(x',y')$ are idempotents of~$S$.
Since~$\sigma(x,y)\gL\sigma(x',y')$,
$\sigma(x,y)=a\sigma(x',y')$ for some~$a\in S$.
Hence, $\sigma(x,y)\sigma(x',y')=a\sigma(x',y')^2=a\sigma(x',y')=\sigma(x,y)$.

Else, if there exists a non-regular $\gH$-class in~$L$.
This means that $L$ contains at least two $\gH$-classes.
Define~$s^L_{\beta,\sigma}(0_\beta)=1$,
and $s^L_{\beta,\sigma}=s^H_{\beta,\sigma'}(x)+1$
elsewhere. The split $s^L$
defined this way is forward ramseyan for~$(\beta,\sigma)$
as above.
It has height at most~$|H|+1\leq 2|H|\leq|L|$.

And this construction satisfies the
determinism property.
\qed
\end{proof}

\subsubsection*{Case of a single~$\gD$-class.}

\begin{lemma}\label{lemma:simon-deterministic-D}
Let~$D$ be a regular~$\gD$-class.
For every ordinal~$\beta$ such that~$\sigma(\beta)\subseteq D$,
there exists a ramseyan split~$s^D_{\beta,\sigma}$
of height at most~$|D|$. Furthermore $s^D$ satisfies the
determinism property.
\end{lemma}
\begin{proof}
We prove the property for every $\gL$-closed $E\subseteq D$.
This is done by induction on the cardinal of~$E$.
If~$E$ is an~$\gL$-class, Lemma~\ref{lemma:simon-determinist-L}
concludes.

Else, let~$L$ be an~$\gL$-class
in~$E$.
Let~$\gamma=\{0_\beta\}\cup\{x\in\dot\beta~:~\sigma(0_\beta,x)\in E\setminus L\}$.
By Fact~\ref{fact:LR-prolong},
for every~$x<y$ in~$\gamma$, $\sigma(x,y)\in E\setminus L$.
On can apply the induction hypothesis,
and obtain a split~$s^{E\setminus L}$ which is forward ramseyan for~$(\gamma,\sigma)$
and of height at most~$|E|-|L|$.
Similarly, for every~$x<y$ in~$\beta\setminus\gamma$,
$\sigma(x,y)\in L$. By Lemma~\ref{lemma:simon-determinist-L},
one obtains a split~$s^L$ which is forward ramseyan
for~$(\beta\setminus\gamma,\sigma)$ of height at most~$|L|$.
Let us define the split~$s^E$ by $s^E(x)=s^{E\setminus L}(x)+|L|$ if~$x\in\gamma$,
else~$s^E(x)=s^L(x)$ if~$x\in\beta\setminus\gamma$.
The mapping~$s^E$ is forward ramseyan for~$(\beta,E)$ as an inheritance
of the forward ramseyanity of~$s^L$ and~$s^{E\setminus L}$.
It has height at most $|E|-|D|+|D|=|E|$.
\qed
\end{proof}

For the proof of Theorem~\ref{theorem:simon-deterministic},
we use Lemma~\ref{lemma:simon-deterministic-D} with~$E=D$,
and the same trick as for ordinal ramseyan splits.

\section{Compaction of additive labellings}
\label{section:compaction}

A labelling maps pairs of elements to a finite set (the semigroup):
it is defined via a finite number of binary predicates.
In this section we show that the use of (forward)
ramseyan factorisations
permits to encode all this information into a finite number
of unary predicates. Furthermore, we show that
the whole additive labelling can be reconstructed from
those unary predicates via first-order formul\ae.
We call this technique \intro{compaction}.

As above, there are two variants to the technique.
One which usable over complete linear orderings
(Section~\ref{subsection:complete-compaction}),
and one usable over ordinals,
which satisfies furthermore the determinisism property
(Section~\ref{subsection:ordinal-compaction}).
In Section~\ref{subsection:result-interpretation},
we apply this technique for proving a new result on
monadic interpretations applied to trees.
And in Section~\ref{subsection:consequences-infinite-structures}
we briefly describe how this result impacts on
the theory of finitely presentable infinite structures.

\subsection{Compactions of additive labelling over complete linear orderings}
\label{subsection:complete-compaction}

We prove here the following statement.
\begin{theorem}\label{theorem:compaction-complete}
For every finite semigroup~$(S,.)$ and~$a$
in~$S$, there exists a first-order formula $\mathbf{labelling}_a(x,y)$
of free variables~$x,y$, which uses the ordering relation~$<$
and unary predicates~$p_1,\dots,p_N$ with~$N=\lceil(6|S|+2)\log_2(|S|)\rceil$
such that the following holds\footnote{We did not try to optimize the value of~$N$.}.

For every complete linear ordering $\alpha$
and additive labelling~$\sigma$ from~$\alpha$
to~$S$, there exists subsets~$X_1,\dots,X_N$ of~$\alpha$ such that
for all~$a$ in~$S$ and~$x<y$ in~$\alpha$:
\begin{align*}
\sigma(x,y)=a\qquad&\text{iff}\qquad (\alpha,X_1,\dots,X_N)\models\mathbf{labelling}_a(x,y)\ ,
\end{align*}
in which for every~$i=1\dots N$, $p_i$ is interpreted as~$X_i$.
\end{theorem}

In this proof, we define first the value of~$X_1,\dots, X_N$,
before giving the formul\ae{}.

Using Theorem~\ref{theorem:simon-split}, one obains a ramseyan split~$s$
for~$(\alpha,\sigma)$ of height at most $3|S|$.
To every element~$x$ in~$\alpha$ and~$k$ with~$1\leq k\leq 3|S|$, we furthermore
attach some pieces of information concerning the value of~$\sigma$.
For every $k$ with~$1\leq k\leq 3|S|$,
there are two such informations, $l_k(x)$ and~$r_k(c)$,
taking value in~$S$,
and corresponding to a compaction of what is happening to
the left of~$x$, and to the right of~$x$ respectively.
We give the definition of~$l_k(x)$, the case of $r_k(x)$
being symmetric.
\begin{align*}
l_k(x)&=\begin{cases}
	\text{any value}&\text{if}~L_k(x)=\emptyset\\
	\sigma(z,x)&\text{if}~L_k(x)~\text{has a maximum}~z\\
	a&\text{else, with}~a~\text{such that}~\forall y\in L_k(x).\,\exists z\in L_k(x).~z>y\wedge\sigma(z,x)=a
	\end{cases}
	\\
	&\text{where}~L_k(x) = \{y<x~:~s(y)=k\}
\end{align*}
Note that a consequence of this definition is that, whenver~$x<y$ are
$k$-neighbours, then~$\sigma(x,y)=l_{s(y)}(y)$.
Finally, it is simple to establish that~$N=\lceil(6|S|+2)\log_2(|S|)\rceil$
bits are sufficient for
coding~$(s(x),l_1(x),\dots,l_{s(x)}(x),r_1(x),\dots,r_{s(x)}(x))$.

We have now to construct first-order formul\ae{} which reconstruct
the value of~$\sigma(x,y)$ for every~$x<y$ in~$\alpha$.
We do not provide the formul\ae{} explicitly, but instead describe
functions which can be easily translated into first-order
logic.
Let us treat first the `ascending case'; i.e. compute~$\sigma(x,y)$
for~$x<y$, $s(x)\leq s(y)$, and~$s(z)\geq s(x)$ for all~$z$ in~$[x,y]$.
\begin{lemma}\label{lemma:correctness-compaction-complete-asc}
For every~$x<y$ in~$\alpha$, if~$s(x)\leq s(y)$ and~$s(z)\geq s(x)$
for all~$z$ in~$[x,y]$, then~$\sigma(x,y)=\mathbf{asc}(x,y)$ with:
\begin{align*}
\mathbf{asc}(x,y)&=
	\begin{cases}
	l_{s(x)}(y)\quad	&\text{if}~s(z)>s(x)~\text{for all}~z\in]x,y[\ ,\\
	l_{s(x)}(z)l_{s(x)}(y)	&\text{else for some}~z\in]x,y[~\text{with}~s(z)=s(x)\ .
	\end{cases}
\end{align*}
\end{lemma}
\begin{proof}
Two cases can happen. If for all~$z$ in~$]x,y[$, $s(z)>s(x)$.
This means that~$[x,y[\cap s^{-1}(s(x))=\{x\}$.
Hence, by definition, $l_{s(x)}(y)=\sigma(x,y)$.

Else, there exists~$x'$ be in~$]x,y[\cap s^{-1}(s(x))$.
By definition of~$l_{s(x)}(y)$,
there exists~$y'$ in~$[x',y[\cap s^{-1}(s(x))$
such that~$l_{s(x)}(y)=\sigma(y',y)$.
Let now~$z$ be the one used in the definition of~$\mathbf{asc}(x,y)$.
By definition of~$l_{s(x)}(z)$,
there exists~$z'$ in~$[x,z[\cap s^{-1}(s(x))$
such that~$\sigma(z',z)=l_{s(x)}(z)$.
Finally using the ramseyanity of~$s$,
we deduce~$\sigma(x,y')=\sigma(z',z)=l_{s(x)}(z)$.
Overall~$\sigma(x,y)=\sigma(x,y')\sigma(y',y)
=l_{s(x)}(z)l_{s(x)}(y)=\mathbf{asc}(x,y)$.\qed
\end{proof}
Naturally, there is a corresponding definition for~$\mathbf{desc}(x,y)$
satisfying~$\sigma(x,y)=\mathbf{desc}(x,y)$ whenever
$s(x)\geq s(y)$ and $s(z)\geq s(y)$ for all~$z$ in~$[x,y]$.
Combining $\mathbf{asc}$ and~$\mathbf{desc}$ we obtain the following.
\begin{lemma}\label{lemma:correctness-compaction-complete}
For every~$x<y$ in~$\alpha$, $\sigma(x,y)=\mathbf{labelling}(x,y)$ with:
\begin{align*}
\mathbf{labelling}(x,y)&=
	\begin{cases}
	\mathbf{asc}(x,y)
	&\text{if}~s(x)\leq s(y)~\text{and}~s(z)\geq s(x)~
		\text{for all}~z~\text{in}~[x,y]\\
	\mathbf{desc}(x,y)
	&\text{if}~s(x)> s(y)~\text{and}~s(z)\geq s(y)~
		\text{for all}~z~\text{in}~[x,y]\\
	\mathbf{desc}(x,z)\mathbf{asc}(z,y)\quad&\text{else, for}~
		z\in]x,y[~\text{and}~s(z')\geq s(z)~\text{for all}~
		z'\in[x,y]\ .
	\end{cases}
\end{align*}
\end{lemma}
\begin{proof}
There are three cases, corresponding to the three items of the definition.
The two first one are treated by
Lemma~\ref{lemma:correctness-compaction-complete-asc} and its variant
for $\mathbf{desc}(x,y)$.
In the third case, one finds~$z$ in~$]x,y[$
such that~$s(z)$ is minimum. We use Lemma~\ref{lemma:correctness-compaction-complete-asc} between~$x$
and~$z$, and its variant
for $\mathbf{desc}$ between~$z$ and~$y$,
as well as the additivity of the labelling~$\sigma$,
for obtaining:
\begin{align*}
\sigma(x,y)&=\sigma(x,z)\sigma(z,y)=\mathbf{asc}(x,z)\mathbf{desc}(z,y)=\mathbf{labelling}(x,y)\ .
\end{align*}
\qed
\end{proof}

It is not difficult at this point to check that
the definition of~$\mathbf{labelling}$
can be translated for every~$a$ in~$S$
into a first-order formula~$\mathbf{labelling}_a$
using as predicate the ordering relation~$<$
as well as unary predicates~$p_1,\dots,p_N$ encoding the value
of~$(s(x),l_1(x),\dots,l_{s(x)}(x),r_1(x),\dots,r_{s(x)}(x))$,
and satisfying the conclusion of Theorem~\ref{theorem:compaction-complete}.

\subsection{Deterministic compaction of additive labellings over ordinals}
\label{subsection:ordinal-compaction}

We now state a result similar to Theorem~\ref{theorem:compaction-complete}
in the ordinal case, which satisfies a form of determinism
property. The statement in itself is difficult to process; it is similar to
the statement of Theorem~\ref{theorem:compaction-complete},
in which the determinism feature has been injected.

\begin{theorem}\label{theorem:compaction-deterministic}
For every finite semigroup~$(S,.)$ and~$a$
in~$S$, there exists a first-order formula $\mathbf{labelling}_a(x,y)$
of free variables~$x,y$, which uses the ordering relation~$<$
and unary predicates~$p_1,\dots,p_N$ with~$N=\lceil(2|S|+1)\log_2(|S|)\rceil$
such that the following holds.
For every ordinal $\alpha$
and additive labelling~$\sigma$ from~$\alpha$
to~$S$, there exists subsets~$X_1(\alpha,\sigma),\dots,X_N(\alpha,\sigma)$
of~$\alpha$ such that
for all~$a$ in~$S$ and~$x<y$ in~$\alpha$:
\begin{align*}
\sigma(x,y)=a\qquad&\text{iff}\qquad (\alpha,X_1(\alpha,\sigma),\dots,X_N(\alpha,\sigma))
	\models\mathbf{labelling}_a(x,y)\ ,
\end{align*}
in which for every~$i=1\dots N$, $p_i$ is interpreted as~$X_i(\alpha,\sigma)$.

Furthermore, for every additive labellings~$\sigma$ and~$\sigma'$
over the respective ordinals~$\alpha$ and~$\alpha'$, and every ordinal
$\beta<\min(\alpha,\alpha')$,
\begin{align*}
\text{if}~\sigma|_\beta=\sigma'|_\beta\quad
&\text{then for all}~i,~~ \beta\in X_i(\alpha,\sigma)~\text{iff}~\beta\in X_i(\alpha,\sigma')
&\text{(determinism property)\ .}
\end{align*}
\end{theorem}

Let~$s$ be the forward ramseyan split of~$(\alpha,\sigma)$ of height~$|S|$
obtained by Theorem~\ref{theorem:simon-deterministic}.
Let us define~$l_k(x)$ as in the previous section (this time only for every~$k=1\dots|S|$).
Without loss of generality, we assume that there exists a neutral element --- denote it $1$ ---
in~$S$, and we set for every~$x$, $\sigma(x,x)=1$. Define:
\begin{align*}
\mathbf{labelling}(x,y)&=\mathbf{labelling}^1(x,y)\ ,
\end{align*}
with $\mathbf{labelling}^n$ defined by induction for all $n=1,\dots,|S|+1$ by:
\begin{align*}
\mathbf{labelling}^n(x,y)&=
	\begin{cases}
	1\quad&\text{if}~n=|S|+1\ ,\\
	\mathbf{labelling}^{n+1}(x,y)&
		\text{else if}~[x,y[\cap s^{-1}(n)=\emptyset\ ,\\
	\mathbf{labelling}^{n+1}(x,z)l_n(y)&
		\text{else if}~[x,y[\cap s^{-1}(n)=\{z\}\ ,\\
	\mathbf{labelling}^{n+1}(x,z_0)l_n(z_1)l_n(y)~&
		\text{else if}~[x,y[\cap s^{-1}(n)=\{z_0<z_1<\dots\}\ .
	\end{cases}
\end{align*}
In this definition,
we abbreviate by $[x,y[\cap s^{-1}(n)=\{z_0<z_1<\dots\}$ the fact that
$z_0$ is the minimal element, and $z_1$ the minimal element different from~$z_0$
in~$[x,y[\cap s^{-1}(n)$. Those two elements exist since $\alpha$ is an ordinal
and since the case of~$[x,y[\cap s^{-1}(n)$ being the emptyset or a singleton
is treated above.

The correctness is then stated by the following lemma.
\begin{lemma}
For every~$x<y$ in~$\alpha$, and~$n=1,\dots,|S|+1$,
if for all~$z$ in~$[x,y[$, $s(z)\geq n$,
then
\begin{align*}
\mathbf{labelling}^{n}(x,y)&=\sigma(x,y)\ .
\end{align*}
\end{lemma}
\begin{proof}
The proof is done by a downward induction on~$n$.
For~$n=|S|+1$, no~$z$ does satisfy~$s(z)\geq n$,
hence $[x,y[$ has to be empty. It follows that $x=y$,
and by consequence~$\mathbf{labelling}^n(x,y)=1=\sigma(x,y)$.

Else, let~$n\leq|S|$.
Assume the property true for~$n+1$ and consider~$x\leq y$.
Let~$E$ be~$[x,y[\cap s^{-1}(n)$.
If~$E$ is empty, this means that for all~$z$ in~$[x,y[$, $s(z)\geq n+1$.
And by induction hypothesis $\mathbf{labelling}^{n+1}(x,y)=\sigma(x,y)$.
Hence, $\mathbf{labelling}^{n}(x,y)=\sigma(x,y)$.
If~$E$ is the singleton~$\{z\}$. This means that~$l_n(y)=\sigma(z,y)$.
It follows that
$\mathbf{labelling}^{n}(x,y)=\mathbf{labelling}^{n+1}(x,z)l_n(y)=\sigma(x,z)\sigma(z,y)=\sigma(x,y)$.
Finaly, if~$E=\{z_0<z_1<\dots\}$. By definition of~$l_n(z_1)$,
$l_n(z_1)=\sigma(z_0,z_1)$. By induction hypothesis,
$\mathbf{labelling}^{n+1}(x,z_0)=\sigma(x,z_0)$.
Furthermore, by definition of~$l_n(y)$,
there is some~$z>z_0$ such that $l_n(y)=\sigma(z,y)$.
Alltogether with the forward ramseyanity of~$s$ leads to:
\begin{align*}
\mathbf{labelling}_n(x,y)
	&=\mathbf{labelling}^{n+1}(x,z_0)l_n(z_1)l_n(y)\\
	&=\sigma(x,z_0)\sigma(z_0,z_1)\sigma(z,y)\\
	&=\sigma(x,z_0)\sigma(z_0,z)\sigma(z,y)\\
	&=\sigma(x,y)\ .
\end{align*}
\qed
\end{proof}

As in the previous case, the construction is easily adaptable
into a presentation by first-order formul\ae{} using
the relation~$<$ together with
$N=\lceil(2|S|+1)\log_2(|S|)\rceil$ unary predicates coding
all the possible values of~$(s(x),l_1(x),\dots,l_{|S|}(x))$.
This concludes the proof of Theorem~\ref{theorem:compaction-deterministic}.

\subsection{Application to interpretations}
\label{subsection:result-interpretation}

We prove in this section Theorem~\ref{theorem:interpretation-decomposition}.
Let us first give two lemmas which are consequences
of standard techniques;
either the compositional method, or tree automata.
\begin{lemma}\label{lemma:multi-variables}
Every monadic formula~$\Phi(x_1,\dots,x_n)$
is equivalent on trees to a formula of the form $\exists z_1\dots\exists z_k.\Phi'$
where~$\Phi'$ is a boolean combination of monadic formul\ae{} of the
form~$x\sqsubset y\wedge \Psi(x,y)$ (of free variables~$x,y$),
$\Psi(x)$ (of free variable $x$)
and~$x=y$, for $x,y$ ranging in
$\{x_1,\dots,x_n,z_1,\dots,z_k\}$.
\end{lemma}

\begin{lemma}\label{lemma:monadic-semigroup}
For every monadic formula of the form~$x\sqsubset y\wedge\Phi(x,y)$ of free variables~$x,y$,
there exists a semigroup~$S_\Phi$ and $A_\Phi\subseteq S_\Phi$
such that, for every tree~$t$, there exists a mapping $\sigma$
which to every nodes~$x\sqsubset y$ associates $\sigma(x,y)\in S_\Phi$, such that
\begin{itemize}
\item $\sigma$ restricted to every branch is an additive mapping, and
\item for every nodes $x\sqsubset y$, $t\models\Phi(x,y)$ iff $\sigma(x,y)\in A_\Phi$.
\end{itemize}
Furthermore, $\sigma$ is monadically definable: for every~$s\in S_\Phi$,
there exists a monadic formula~$\Phi_s(x,y)$ such that for every tree~$t$
and nodes~$x\sqsubset y$, $t\models\Phi_s(x,y)$ iff~$\sigma(x,y)=s$.
\end{lemma}

And the result is then the following.
\begin{theorem}\label{theorem:interpretation-decomposition}
For every monadic interpretation~$\interpretation_{\mathit{MSO}}$,
there exists a monadic marking $\mathcal{M}_{\mathit{MSO}}$
and a first-order interpretation~$\interpretation_\mathit{FO}$
such that for every tree~$t$, $\interpretation_{\mathit{MSO}}(t)=\interpretation_\mathit{FO}(\mathcal{M}_{\mathit{MSO}}(t))$.
\end{theorem}
\begin{proof}
Wlog, we prove the result for an interpretation~$\interpretation_{\mathit{MSO}}$
with a single formula $\Phi(x_1,\dots,x_n)$.
Using Lemma~\ref{lemma:multi-variables},
we just have to show how to obtain an equivalent to a formula
of the form~$x\sqsubset y\wedge\Psi(x,y)$ as the combination of
a monadic marking and a first-order formula.
For this, we use Lemma~\ref{lemma:monadic-semigroup}
which tells us that the value of~$\Psi(x,y)$ can be uncovered
by projection of an additive labelling.
And we use Theorem~\ref{theorem:compaction-deterministic}
for reducing the computation of the additive labelling
to the combination of a monadic marking and a first-order formula.

Note that this argument heavily relies on the determinism
of the construction of Theorem~\ref{theorem:compaction-deterministic}.
Indeed, one has to mark every branch of a tree, \emph{a priori} with a different
marking. The determinism property allows to have a single marking
for the whole tree.
\qed
\end{proof}

\subsection{Consequences for infinite structures}
\label{subsection:consequences-infinite-structures}

The goal of this section is to show how the results given above,
namely Theorem~\ref{theorem:interpretation-decomposition},
have direct new consequences in the definition of some families of
finitely presentable infinite structures.
There is no real technical contribution in this section but rather a presentation
of those consequences to the theory of infinite structures.
Let us warn the reader that
we do not intend to provide a survey of this area, since this would require
much more space and would be out of topic.
We rather directly concentrate on providing Theorems~\ref{theorem:carac-pr}
and~\ref{theorem:carac-cauc}. Essentially, those results show that for
the standard caracterisation of the families of prefix-recognizable graphs,
as well as for the Caucal hierarchy, one can replace the monadic interpretations
by first-order ones.

The prefix-recognizable graphs were introduced by Caucal via an internal definition \cite{caucal96}.
Namely, fix a finite alphabet~$A$.
A \intro{prefix-recognizable} graph is an infinite directed graph defined as follows.
Its set of vertices is a regular language over the alphabet~$A$.
And each edge relation is a finite union of relations of the form~$(U\times V).W$ with
\begin{align*}
(U\times V).W &= \{(uw,vw)\,:\,u\in U,\,v\in V,\,w\in W\}\ ,
\end{align*}
for~$U,V,W$ regular languages.
By extension, a graph is \intro{prefix recognizable} if it is isomorphic to such a graph.
An important property of those graphs is that their monadic theory is decidable (this fact is due
to Caucal \cite{caucal96}; it can be easily seen
as a direct consequence of Rabin Theorem~\cite{rabin69} stating that the complete binary tree
has a decidable monadic theory, together with Theorem~\ref{theorem:carac-pr-mso} below).

There exists different caracterisations for this class of graphs.
We will use below the following one:
\begin{theorem}[Blumensath \cite{blumensath01}]\label{theorem:carac-pr-mso}
A graph is prefix-recognizable iff it is isomorphic to a monadic interpretation of
the complete binary tree.
\end{theorem}
Using this theorem as guide, one can extend the definition of prefix-recognisability to
relational structures:
we call a relational structure \intro{prefix-recognizable}
if it is monadically interpretable in the complete binary tree.

Theorem~\ref{theorem:interpretation-decomposition} provides another --- new ---
caracterisation of prefix-recognizable structures, Theorem~\ref{theorem:carac-pr}.
Beforehand, we need the following lemma.
\begin{lemma}\label{lemma:reg-fo-int}
Let~$t$ be a regular tree.
Then there exists a first-order interpretation $\IFO$ such that~$t$ is isomorphic to
$\IFO(\cbt)$.
\end{lemma}
\begin{proof}
It is sufficient to consider that the regular tree is the complete
binary tree together with a regular labelling in some finite alphabet~$A$
attached to  every node.
This means that there exists a deterministic
and complete finite automata $\mathcal{A}$
of finite words over the alphabet~$\{0,1\}$, with each state labelled by a letter in~$A$,
such that the label
of a node~$u$ is the letter attached to the sole state reached from the initial state
while reading~$u$.
Let this automaton have states~$Q$, initial state~$q_0$, and transition function $\delta$
from~$Q\times\{0,1\}$ to~$Q$. As usual we extend this transition function into
a mapping from~$Q\times\{0,1\}^*$ to~$Q$.
Wlog we can assume that there exists also a mapping $d$ from $Q$ to~$\{0,1\}$
such that for every state~$q$ in~$Q$ and letter~$a$ in~$\{0,1\}$,
$d(\delta(q,a))=a$; i.e. the automaton remembers whether the current node
is a left or a right child.

Let~$n$ be a mapping numbering the states of~$\mathcal{A}$ from~$1$ to $|Q|$.
Given a word~$u=a_1a_2\dots a_n$, the $a_i$'s being letters in~$\{0,1\}$,
define:
\begin{align*}
f(u)&=10^{n(q_0)}10^{n(q_1)}1\dots10^{n(q_n)}1
\end{align*}
in which $q_0,q_1,\dots,q_n$ are the $n+1$ states successively assumed by the automaton
while reading the word~$u$. The proviso concerning
the mapping~$d$ makes $f$ an injective mapping.

The image of~$f$ is first-order definable (as a language of words).
Indeed, in order to verify that a word belongs to the image of~$f$,
it is sufficient to check a) that $10^{n(q_0)}1$ is a prefix, b) that the last letter is~$1$,
and c) that every factor of the form $10^n10^m1$ is such that~$n=n(p)$ and~$m=n(q)$
for some transition~$\delta(p,a)=q$. Those verifications are first-order definable.
Furthermore, for every word~$u$, the state $\delta(q_0,u)$
is nothing but the sole state~$q$ such that $10^{n(q)}1$
is suffix of~$f(u)$. This is also first-order definable.

From those remarks, it is easy to give a first-order
interpretation which, given the complete binary tree,
selects the nodes belonging to the image of~$f$, and labels
every node~$f(u)$ by the state~$\delta(q_0,u)$.
This interpretation provides a new tree~$t'$.
Since all the relevant information --- the label of the node,
and its right-child/left-child nature --- is encoded in each state,
it is easy to first-order interpret~$t$ in~$t'$.
\qed
\end{proof}

\begin{theorem}\label{theorem:carac-pr}
A structure is  prefix-recognizable iff it is isomorphic
to the first-order interpretation (with ancestor relation)
of the complete binary tree.
\end{theorem}
\begin{proof}
We have to show that given a monadic interpretation $\IMSO$, there exists
a first-order interpretation~$\IFO$ such that~$\IMSO(\cbt)$
is isomorphic to~$\IFO(\cbt)$.
Using Theorem~\ref{theorem:interpretation-decomposition},
we have that~$\IMSO(\cbt)$ is equal to~$\IFO'(\LMSO(\cbt))$
for some monadic labelling~$\LMSO$ and first-order interpretation $\IFO'$.
Then using Lemma~\ref{lemma:reg-fo-int}, we obtain an interpretation $\IFO''$
such that~$\IFO''(\cbt)$ is isomorphic to~$\LMSO(\cbt)$.
By closure of first-order interpretation under composition, $\IFO=\IFO'\circ\IFO''$
is a first-order interpretation such that~$\IFO(\cbt)$ is isomorphic to~$\IMSO(\cbt)$.
\qed
\end{proof}

A similar approach can be used for caracterising the Caucal hierarchy.
The Caucal hierarchy \cite{caucal02} is an extension of prefix-recognizable graphs
to `higher-order'. We use here the caracterisation of
 Carayol and W\"ohrle \cite{carayolwohrle03} as a definition:
\begin{itemize}
\item The structures in~$\mathit{Struct}_0$ are the finite relational structures.
\item The graphs in~$\mathit{Graph}_n$ are the structures in $\mathit{Struct}_{n}$ having a graph signature.
\item The trees in~$\mathit{Tree}_{n+1}$ are the unfolding of graphs in $\mathit{Graph}_n$.
\item The structures in~$\mathit{Struct}_{n+1}$  are the monadic interpretations of trees in $\mathit{Tree}_{n+1}$.
\end{itemize}
Since both the monadic interpretation and the unfolding preserve the decidability
of the monadic theory,
the trees, graphs and structures in the classes defined above have
a decidable monadic theory.

The following interpretation shows that in the definition of this hierarchy, the monadic logic can be replaced
by first-order logic.
\begin{theorem}\label{theorem:carac-cauc}
The structures in~$\mathit{Struct}_n$  are, up to isomorphism, the first-order interpretation of trees in $\mathit{Tree}_{n}$.
\end{theorem}
In fact, this is a direct consequence of Theorem~\ref{theorem:interpretation-decomposition}
together with the following proposition (see \cite{carayolwohrle03},
Proposition~1).
\begin{proposition}
The class $\mathit{Tree_n}$ is closed under monadic markings.
\end{proposition}

\section*{Acknowledgement}

I am deeply grateful to Achim Blumensath
and Olivier Carton for their help in the
production of this document.

\bibliographystyle{plain}
\bibliography{biblio.bib}

\end{document}